\documentclass[useAMS]{mn2e}

\usepackage{amsmath}
\usepackage{url}
\usepackage{amsfonts}
\usepackage{amsbsy}
\usepackage{verbatim}
\usepackage{array}
\usepackage{amssymb}
\usepackage{graphicx,color,subfig}

\renewcommand{\d}{\ensuremath{\partial}}

\newcommand{\kk}{\ensuremath{\mathbf{k}}}

\title[Thermal hysteresis in rings]{Thermal hysteresis and front propagation in dense planetary rings}
\author[Larue, Latter, Rein]{R\'emy Larue$^{1,2,3}$,
   Henrik Latter$^{1}$\thanks{E-mail:
    hl278@cam.ac.uk},
   Hanno Rein$^{4,5}$ \\
$^{1}$DAMTP, University of Cambridge, CMS, Wilberforce Road,
Cambridge CB3 0WA, UK\\
$^{2}${ENS Paris-Saclay, 4 avenue des Sciences
91190 Gif-sur-Yvette, France}\\
$^{3}${Laboratoire de Physique Subatomique et de Cosmologie, Universit\'{e} Grenoble-Alpes, CNRS/IN2P3, Grenoble INP, 38000 Grenoble, France}\\
$^{4}${Department of Physical and Environmental Sciences, University of Toronto at Scarborough, Toronto, Ontario M1C 1A4, Canada}\\
$^{5}${David A. Dunlap Department of Astronomy and Astrophysics, University of Toronto, Toronto, Ontario, M5S 3H4, Canada}
}

\date{}

\begin{document}

\maketitle

\begin{abstract}

Saturn's rings are composed of icy grains, most in the mm to m size ranges, undergoing several collisions per orbit. Their collective behaviour generates a remarkable array of structure over many orders of magnitude, much of it not well understood. On the other hand, the collisional properties and parameters of individual ring particles are poorly constrained; usually $N$-body simulations and kinetic theory employ hard-sphere models with a coefficient of restitution $\epsilon$ that is constant or a decreasing function of impact speed. Due to plastic deformation of surface regolith, however, it is likely that $\epsilon$ will be more complicated, at the very least a non-monotonic function. We undertake $N$-body simulations with the REBOUND code with non-monotonic $\epsilon$ laws to approximate surfaces that are friable but not sticking. Our simulations reveal that such ring models can support two thermally stable steady states for the same (dynamical) optical depth: a cold and a warm state. If the ring breaks up into radial bands of one or the other state, we find that  warmer states tend to migrate into the colder states via a coherent travelling front. We also find stationary `viscous' fronts, which connect states of different optical depth, but the same angular momentum flux. We discuss these preliminary results and speculate on their implications for structure formation in Saturn's B and C-rings, especially with respect to structures that appear in Cassini images but not in occultations. 
\end{abstract}


\section{Introduction}

Saturn's rings flaunt an extraordinary array of axisymmetric structure,
both quasi-regular and chaotic, ranging over some four orders of magnitude in length - from 10 m to 100 km (Colwell et al.~2009, Cuzzi et al.~2018). Yet despite several decades of theoretical effort, their origins are only partially understood (Schmidt et al.~2009, Estrada et al.~2018, Salo et al.~2018).
In particular, the disjunct bands of high and low optical depth in the B-ring (Horn and Cuzzi 1996, Colwell et al. 2007), the plateaus in the C-ring (Tiscareno et al.~2019), and the irregular intermediate scale striations in the A and B-rings (Porco et al.~2005) are presently without plausible explanations. 
Simply put, there is too much observed structure and too few suitable instabilities (or related processes) in our theoretical models. Perhaps it is time to re-assess some of our fundamental assumptions and explore a wider range of alternative scenarios.   

It is probable, though not assured, that much of the ring's unexplained structure arises spontaneously due to its peculiar granular flow. Since the 1980s researchers have turned to kinetic theory or $N$-body simulations to model this flow, initially calculating the thermal balances underlying ring equilibria, and then the (viscous) instabilities that might generate structure (e.g., H\"ameen-Anttila 1982, Araki \& Tremaine 1986, Wisdom \& Tremaine 1988, Salo 1991, H\"ameen-Anttila \& Salo 1993, Salo et al.~2001, Latter \& Ogilve 2006, 2008). These studies have made several strong assumptions, especially regarding the nature of the ring particles and their collisional behaviour, for instance rarely deviating from a hard-sphere model with either a constant coefficient of restitution $\epsilon$ or a `Bridges law' (Bridges et al.~1984), whereby collisions below some critical impact speed are perfectly elastic.
In reality, ring particles are likely to be irregularly shaped and coated in a regolith of small particles $\lesssim 1$ cm (e.g. Doyle et al.1989, Nicholson et al.~2008, Morishima et al.~2012; Deau 2015) and, being irregular and fluffy, their surfaces should produce an \emph{enhanced inelasticity} at low impact speeds, and indeed possible particle adhesion.  
In light of this, the adoption of a constant $\epsilon$, or a Bridges law, may significantly misrepresent some of the ring's collective collisional dynamics.
Our paper tests this idea by exploring other, physically motivated, prescriptions for $\epsilon$.
We find, in fact, that even very simple changes to the collision law can give remarkably different outcomes.


Continuum mechanical models of viscoelastic collisions that account for fluffy and/or sticky surfaces demonstrate that $\epsilon$ is a non-monotonic function of impact speed $v_\text{coll}$. 
Beneath some critical speed we have $\epsilon=0$, but on increasing $v_\text{coll}$, $\epsilon$ rises, plateaus, and then decreases again (Gorkavyi 1985, Hertzsch 2002, Albers \& Spahn 2006, Brilliantov et al.~2007). 
Laboratory experiments appear to confirm this picture (Gorkavyi 1989, Hatzes et al.~1991, Bridges et al.~1996).
We implement collision laws of this basic form in our paper and term them `regolith laws'. In addition, at or below the critical speed colliding particles may stick, but we neglect this important effect in order to avoid the vexed and complicated issue of size-distribution dynamics (e.g. Brilliantov et al.~2015). 
Our approach is mainly numerical, via $N$-body simulations of monodisperse, spherical, indestructible particles with the code REBOUND; but we also employ a dense gas kinetic theory, where appropriate. 
Note that we do not include self-gravity and thus our simulations fail to exhibit wakes, nor do they support viscous overstability, both important phenomena we hope to test in the future. Our study is distinct but complementary to recent $N$-body simulations that explicitly test the role of adhesion, especially on instabilities (Ballouz et al.~2017, Lu et al.~2018; see also Section 16.7.1.7 in Salo et al.~2018). Our main focus, in contrast, will be on disk thermodynamics.


Our first main result is that regolith laws permit a dense ring to fall into one of two thermally stable states at the same optical depth: (a) a very dense state with filling factors $\sim 0.3$ and low temperatures, $c\lesssim a\Omega$ (where $c$ is velocity dispersion, $a$ is particle radius, and $\Omega$ is orbital frequency) and (b) a moderately dense state  with lower filling factors ($\lesssim 0.1$) and a slightly warmer temperature, $c\gtrsim 4a\Omega$. This bistability generally favours optical depths less than 1, but can be pushed up to higher values if we broaden our parameter range. We also find in certain circumstance that the cold state at low optical depth is metastable: shot noise permits the ring to spontaneously jump into the hot state.


Our second set of results explores what happens when different thermal states spatially adjoin. If two states of the same optical depth but different temperature connect, a travelling `thermal front' develops that can reach speeds of $\lesssim a\Omega$, while maintaining a steady spatial structure. If the front is too slow, the disparity in the angular momentum flux between the two states reorganises the front profile so that the flux is uniform but the optical depth undergoes a jump, what we term a static `viscous front'. Some of the latter behavior mirrors that witnessed by Salo and Schmidt (2010) in their simulations of viscous instability. 


The plan of the paper is as follows. The next section begins with a review of the extant literature on low-impact collisions between regolith covered and/or sticky particles, moving on to a presentation of the model collision laws we use, and then our numerical methods. Subsequently, we detail out results: the calculation of thermal equilibria and hysteresis in smallish boxes (Section 3), potential metastability (Section 4), and finally results on spatially adjoining states, i.e. thermal and viscous front (Section 5). We conclude in Section 6.

\section{Background and methods}

This section presents the physical set-up and numerical model by which we attack the thermal equilibria of rings composed of regolith-coated particles. We first devote some space to set the scene, by reviewing the theoretical and experimental literature and explaining the key ideas and parameters that underlie work in this area. The model collision laws we adopt are then exhibited, followed by the details of the $N$-body simulations with REBOUND we conduct. 

\subsection{Collisional physics and the coefficient of restitution}

We aim to describe the collisional dynamics of many ring particles in a local patch of a planetary ring. From the outset we make several strong assumptions that we concede may distort our results: the particles are taken to be identical, spherical, and frictionless. Most of the ring mass is in metre-sized particles, and thus it is that population that we track.   
Only binary collisions are considered, and these are deemed inelastic, so that $\mathbf{g}'\cdot\kk= -\epsilon(\mathbf{g}\cdot\kk)$, where $\mathbf{g}$ is the relative velocity of two colliding particles before the collision and $\mathbf{g}'$ afterwards, $\kk$ is the unit vector pointing between the two particles centres at the moment of collision, and $\epsilon$ is the coefficient of restitution. This coefficient lies between 0 and 1 and is usually a function of the impact speed $v_\text{coll}=|\mathbf{g}\cdot\kk|$. We neglect the possibility of two particles sticking and assume that all the specifics of the particle surfaces can be encapsulated in the functional behaviour of $\epsilon$. Because we find the ring dynamics are so sensitive to $\epsilon$, we now spend some time discussing this important physical input.

\subsubsection{Theoretical and experimental background}


 Research exploring the collisional behaviour of regolith-covered particles can be separated into analytical calculations, drawing on continuum mechanics, and laboratory experiments, approximating Saturnian conditions. We attempt to review and synthesise this body of work.

The seminal experiments in this area were described in Bridges et al. (1984) and collided smooth ice spheres with an ice block at temperatures $\sim 170$K. This work produced the collision law $\epsilon=\text{min}\left[1,\,(v_\text{coll}/v_\text{crit})^{-0.234}\right]$, for $v_\text{crit}=0.008$ cm s$^{-1}$, a defining feature of which is perfect elasticity at sufficiently low collision speeds ($v_\text{coll}<v_\text{crit}$). This collision law became the standard for subsequent $N$-body simulations and other theoretical work. Subsequently, broken power laws of this type were shown to arise naturally in generalisations of the Hertz theory to viscoelastic solids (Dilley 1993, Hertzsch et al. 1995, Brilliantov et al. 1996, Thornton 1997). However, such theoretical work must posit that the surfaces of the colliding spheres are smooth and that irreversible energy losses arise solely from viscoelastic deformations inside the spheres.

Shortly after the Bridges experiments, two neglected but insightful papers by Gorkavyi (1985, 1989) highlighted the importance of regolith and argued against perfectly elastic restitution at low impact speed. Gorkavyi emphasised that $\epsilon$ can be dramatically altered at small $v_\text{coll}$ because (a) impact energy can be used up when reshaping a soft friable surface (leaving nothing left over for elastic rebound) and/or (b) rebounding motion can be countered by surface stickiness.
Using energy arguments, the 1985 paper sketches out three regimes: (a) at sufficiently low $v_\text{coll}$, there is total energy loss and thus $\epsilon=0$ (sticking/adhesion is not considered); (b) at slightly larger $v_\text{coll}$, $\epsilon$ increases with $v_\text{coll}$; and then (c) after a turning point, $\epsilon$ decreases with $v_\text{coll}$ (traditional restitution). The collision law is hence non-monotonic. Gorkavyi (1989) followed this up with simple experiments using powders, metals, and marble at room temperature and pressure, which agree with earlier lab work by Hartmann (1978, 1985), in a different context, using rocks.

Subsequent papers from the Bridges research group examined how the state of the particle surface influenced collisions, with a particular focus on the adhesive effect of frost, a thin layer of microscopic structure that might behave similarly to the thicker regolith layer expected on larger ring particles. Hatzes et al.~(1991) showed frosty particles can stick at impact speeds below some critical level (a few mm s$^{-1}$), but did not examine explicitly how it changed the form of $\epsilon$. Bridges et al.~(1996) conducted a large set of experiments for different kinds of ices and $v_\text{coll}$ at relevant temperatures, which further strengthened the case for sticking, and also showed that $\epsilon$ exhibited the three main features predicted by Gorkavyi.


On the theoretical side, the 2000s witnessed various extensions of Hertz contact mechanics, accounting for both viscoelasticity and particle adhesion via JRK theory (Albers and Spahn 2006, Brilliantov et al. 2007; see also Thornton and Ning 1998, and Chokshi et al.~1993, the latter in the context of ISM grains). Notable is the work by Hertzsch (2002) who modelled the two effects of sticking and of passive regolith deformation, as discussed by Gorkavyi, treating the passive regolith as a deformable viscous non-sticky `soft layer'. Both physical effects appear to influence the form of $\epsilon$ similarly. In all cases, non-monotonic $\epsilon$ laws were mathematically derived.

Brilliantov et al.~(2007) provides estimates for solid water-ice particles of various sizes that, despite several strong assumptions, help with Saturnian applications.
For metre-sized water-ice impactors, the theory predicts that the maximum value $\epsilon$ takes is relatively large, potentially above 0.7. For cm sized particles, it drops to $\approx 0.3$. On the other hand, the critical $v_\text{coll}$ for sticking is roughly $10^{-2}$ cm s$^{-1}$ for metre-sized ice impactors, and this rises to greater than 0.1 cm s$^{-1}$ for cm-sized particles. Because of the model assumptions care must be taken, however, when applying these estimates, and in fact the quoted critical collision speeds are probably gross lower limits. The theory omits the energy dissipation channel associated with irreversible regolith deformation (as well as internal fracture) by treating the particles as solid-ice non-spinning viscoelastic spheres. It also sets the unknown dissipative constant $A$ by fitting a (non-sticking) viscoelastic model (Brilliantov et al.~1996) to the (non-sticking) experimental data of Bridges et al. Nonetheless, the Brilliantov results provide a useful starting point for our study. 

Before moving on, we flag additional physics not yet discussed. In applying the above ideas and prescriptions to an ensemble of colliding particles, one must acknowledge that, by virtue of the collisions themselves, particles' surface properties will evolve. Repeated collisions will presumably `compactify'  particle regolith and hence reduce the mean critical sticking speed. On the other hand, bombardment by micrometeoroids will disturb the surfaces and there will be accretion of very small floating particles, processes that will rejuvenate regolith.
It follows that, in addition to the size distribution dynamics (e.g. Longaretti 1989, Bodrova et al. 2012, Brilliantov et al.~2015), there will take place related dynamics controlling the mean surface properties. We do not attempt to construct a model for this interesting process here.





\subsubsection{Important scales}

This subsection briefly outlines the key velocity scales relevant for our problem. We assume that there is a single critical sticking speed $v_\text{stick}$ below which two impactors will adhere. We also assume a second critical impact speed $v_\text{crit}$ below which $\epsilon=0$. It may be that these two speeds are the same, though in general we expect $v_\text{stick}< v_\text{crit}$, i.e. it is possible for all the energy of the impact to be used up reshaping the surface and resisting the adhesive attraction of the regolith, thereby allowing the impactors to roll clear of each other. Particle spin and tidal shear may facilitate such non-sticking $\epsilon=0$ encounters. 

A third key speed is the velocity dispersion $c$, as impact speeds will be distributed around it. Thus the relative size of $c$ relative to $v_\text{crit}$ will determine which collisional regime (sticking, non-sticking, etc.) the particles are in. Partly controlling $c$ is the orbital shear speed across a particle, $a\Omega$ (recall $a$ is particle radius and $\Omega$ the orbital frequency). The importance of this scale issues from the fact that dense cold rings adopt a velocity dispersion $ c \sim a\Omega$, in the absence of gravity wakes, and $c\lesssim 5 a\Omega$, when gravity wakes are present (e.g., Araki \& Tremaine 1986, Salo et al.~2018)\footnote{The second estimate can be obtained by assuming a gravitationally unstable ring settles into a state where the Toomre $Q$ is $\sim 1$, and then taking typical values for the surface density (e.g. Hedman \& Nicholson 2013, 2016)}. It follows that if $c\sim a\Omega\gg v_\text{crit}$ then the regolith is not going to feature much in the mean thermal dynamics, and hence the determination of $c$. On the other hand, if $c\sim a\Omega \ll v_\text{crit}$ then the surface properties are going to be important. Complicating this picture, of course, is the size dependence of both $a\Omega$ and $v_\text{crit}$. In a polydisperse ring, however, the velocity dispersion of smaller particles will be similar to the metre-sized particles (Salo et al.~2018). We now obtain some bounds on the important parameter $v_\text{crit}/(a\Omega)$.

First we situate ourselves at a representative location in the C-ring, in which gravity wakes are likely absent, and set $\Omega\approx 10^{-4}$ s$^{-1}$. If $a=1$ m, the most dynamically important size, $a\Omega$ is roughly~0.01 cm/s. Next, applying the estimates from Brilliantov et al.~(2007) (cf. Section 2.1.1) and setting $v_\text{crit}=v_\text{stick}$, we obtain $v_\text{crit}/(a\Omega)\sim 1$. 
For cm sizes, $v_\text{crit}/(a\Omega)\gtrsim 10$ (noting that the velocity dispersion of this population is set by the metre sizes). As argued earlier, the Brilliantov estimates for $v_\text{crit}$ only provide lower bounds, and hence we conclude that it is likely that the C-ring is in a regime where surface regolith properties will matter.

At a representative location in the A or B-ring, we must take into account gravity wakes. Thus we find ourselves in a more ambiguous situation: the Brilliantov estimates yield $v_\text{crit}/c\gtrsim 0.1$ for metre-sized particles, and $v_\text{crit}/c \gtrsim 1$ for cm-sized particles. Depending on how badly the Brilliantov results underestimate $v_\text{crit}$, we could be in a marginal regime or in a regolith-dominated regime. Certainly, further work on the collisional dynamics of ice would help decide on this point. As we do not simulate self-gravity, for now we just assume that $a\Omega < v_\text{crit}$, and leave open its importance to future work.

\subsubsection{Model coefficients of restitution}

This section presents the two classes of  non-monotonic `regolith' $\epsilon$-law we use in this paper. We have attempted to paramaterise these laws in two readily understandable quantities: $v_\text{crit}$, the impact speed at which collisions are perfectly inelastic (cf. Section 2.1.2); and $\epsilon_\text{max}$, the turning point value of $\epsilon$ (i.e., its maximum).



\begin{figure}
    \centering
    \includegraphics[width=0.5\textwidth]{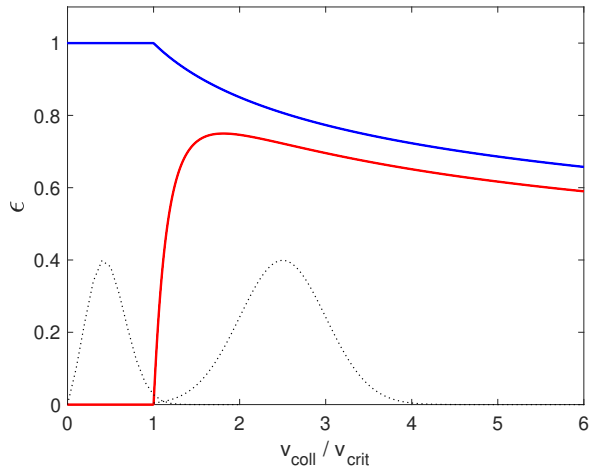}
    \caption{Two forms of the coefficient of restitution $\epsilon$ as a function of impact speed $v_\text{coll}$. The solid blue curve is the Bridges law, see Eq.~\eqref{BPL}, with $\epsilon_0=1$. The red solid curve is the `regolith' law, Eq.~\eqref{regolith}, with $b=(1/4)v_\text{crit}$ and $\epsilon_\text{max}=0.75$. In addition, we have sketched two velocity distribution functions with black dotted curves; see discussion in Section \ref{bistability}. }
    \label{epsilons}
\end{figure}

A broken power law (BPL) for $\epsilon$, though somewhat crude has the benefits that it has few input parameters and some headway can be made with it using kinetic theory. We define the law in the following way:
\begin{equation} \label{BPL}
    \epsilon (v_{\mathrm{coll}})=\begin{cases}
    \epsilon_0, & \text{if $v_{\mathrm{coll}}<v_{\mathrm{crit}}$}.\\
    \epsilon_\text{max}\left(v_{\mathrm{coll}} / v_\text{crit}\right)^{-p}, & \text{if $v_{\mathrm{coll}}\geq v_{\mathrm{crit}}$}.
  \end{cases}
\end{equation}
We set the exponent $p=0.234$, following Bridges et al.~(1984), though it could take other values. The quantity $\epsilon_0$ we set equal to either 1, to obtain the Bridges et al.\ law itself, or equal to 0, to get the opposite perfectly inelastic law. The Bridges BPL is plotted in Fig.~\ref{epsilons} in blue.


A more realistic non-monotonic $\epsilon$ law that is smoother and exhibits something of a plateau near its maximum can be defined in several ways. We choose the following:
\begin{equation} \label{regolith}
    \epsilon (v_{\mathrm{coll}})=\begin{cases}
    0, & \text{if $v_{\mathrm{coll}}<v_{\mathrm{crit}}$}\\
     1.625\,\epsilon_{\mathrm{max}}\,\zeta/(1+\zeta^{1.234}), & \text{$v_{\mathrm{coll}}\geq v_{\mathrm{crit}}$},
  \end{cases}
\end{equation}
where $\zeta=(v_{\mathrm{coll}}-v_{\mathrm{crit}})/{b}$ and 
$b$ is the plateau `width', usually set to $a\Omega$.
 Constants have been chosen so that $\epsilon$ approaches the Bridges law for large $v_{\mathrm{coll}}$. To facilitate the discussion later, when we compare the different models, we refer to Eq.~\eqref{regolith} as a `realistic' law (though it is yet to be determined how realistic it is). We plot it in Fig.~\ref{epsilons} in red. 

\subsection{The potential for bistability}
\label{bistability}

Before presenting our numerical methods and the results that ensue, we briefly explain why a non-monotonic collision law, such as given in Eq.~\eqref{regolith} and displayed in Fig.~\ref{epsilons}, potentially yields two stable states for the same parameters.   

 At lower optical depths, $N$-body simulations and kinetic theory show that the Bridges law yields equilibria with $c > a\Omega$, and thus most collisions sample the power-law decreasing segment of the $\epsilon$ curve (Salo 1991, Latter \& Ogilvie 2008). As mentioned above, the realistic regolith law we adopt approaches the Bridges law for impact speeds larger than the turning point in $\epsilon$, and is a reasonable approximation near the turning point. One might then expect that collisions employing the regolith law would sample similar values of $\epsilon$ and the resulting thermal equilibria will resemble the Bridges equilibria, giving us a `warm' ring. In Fig.~\ref{epsilons} we superimpose a mock impact velocity distribution at larger $v_\text{coll}$ to indicate such a state. 
 
 On the other hand, when $\epsilon$ is a constant and taken to be equal to zero the thermal equilibria are especially cold, with $c\sim a \Omega$ (e.g. Araki \& Tremaine 1986). It follows that our regolith law might be capable of supporting these very cold equilibria as well. This should certainly be the case if $v_\text{crit}$ is much larger than $a\Omega$. In this circumstance, most impact speeds will fall below $v_\text{crit}$ and thus yield perfectly inelastic collisions with $\epsilon=0$, never sampling the non-zero segment of the $\epsilon$ curve. Fig.~\ref{epsilons} indicates a schematic velocity distribution for this state, centred on a value less than $v_\text{crit}$.
 
 Both the warm state and the cold state are thermally stable, as has been shown separately in $N$-body simulations.
 \emph{And thus a non-monotonic law may yield bistability.}
 The disk may fall into either the cold or the warm homogeneous state for exactly the same parameters (most notably optical depth $\tau$)\footnote{This bistability is different to the `phase transitions' associated with viscous instability, which drives the system to a \textit{non-homogeneous} state characterised by abutting radial regions of high and low optical depth (e.g., Lukkari 1981, H\"ameen-Anttila 1982, Salo \& Schmidt 2010).}. Which is chosen depends on the initial conditions. Moreover, it follows there must also be an intermediate thermally unstable state separating the two stable states, though this will not normally be observed. The argument for bistability is strongest in a regime where $v_\text{crit}\gg a\Omega$. A question then is: what is the minimum value of $v_\text{crit}$ that yields bistability? Our simulations results in Section 3 aim to answer this and other questions.

\subsection{N-body simulations} 

In this subsection we further outline  the physical model we adopt and the numerical methods used to calculate its non-trivial thermal dynamics. We seek to determine the evolution of a large number of inelastically colliding particles, and thus our main tool will be local $N$-body simulations.   

\subsubsection{Equations of motion}

We solve the equations of motion in the Hill approximation (Hill 1878), a local coordinate system that is co-rotating with a particle on a circular orbit. 
The gravity from the central object is linearized in local coordinates and the orbital frequency is a constant. 
This allows, but does not restrict, us to use shear-periodic boundary conditions.
In that case, the Hill approximation is also referred to as the shearing sheet. 
In our notation, the $x$, $y$, and $z$ coordinates point in the radial, azimuthal and vertical direction, respectively. 

Treating the central object, Saturn, as a point source, the equations of motion for a test particle can be written as
\begin{align}
\ddot{x} &=2\Omega \dot{y} + 3\Omega^2 x +F_x^\text{coll}, \\ 
\ddot{y} &= -2\Omega \dot{x}+F_y^\text{coll}, \\
\ddot{z} &= \Omega^2 z+F_z^\text{coll},
\end{align}
where $\mathbf{F}^\text{coll}$ is the (intermittent) acceleration exerted on a particle during a collision. 
In the absence of collisions, the solution to these equations can be written as epicycles (e.g. Rein \& Tremaine 2011). 

The particles move within a finite-size numerical domain/box. We denote the radial length of the box by $L_x$ and the azimuthal length by $L_y$. 
In all our experiments, the vertical length of the box $L_z$ has been chosen to be large enough so that no particle ever crosses the vertical boundaries. Otherwise, the box is periodic in $y$ and shear-periodic in $x$. 

The only further ingredients needed are the finite particle radius $a$ and a collision model. 
We treat particles as hard spheres (they are not permitted to overlap) and the outcome of a collision is described using a normal coefficient of restitution, as described in Section 2.1.3. 
The particles have no spin.

\subsubsection{Numerical method}

We use the freely available $N$-body code REBOUND (Rein \& Liu 2012) to perform all of the simulations presented in the paper. 
To evolve the equations of motion forward in time, we use the Symplectic Epicycle Integrator (SEI, Rein \& Tremaine 2011) which is well suited for simulations of particle motion within the Hill approximation.

Collisions are detected using a nearest neighbour tree search. 
We randomize the order in which collisions are resolved after each timestep. 
We found that this removes spurious correlations which might otherwise be introduced when choosing a specific order in which collisions are resolved (i.e. resolving them from left to right, by a numerical particle identifier, or by the position in memory).

\subsubsection{Diagnostics}

In order to probe the collective behaviour of the granular flow, we require a number of averaged quantities. We define the mean normal geometrical optical depth $\tau$ as the total projected area of the particles on the $(x, y)$ plane divided by the total area of the $(x, y)$ plane. In other words,
\begin{equation}
\tau =N\pi a^2/(L_x L_y), 
\end{equation}
where $N$ is the number of particles. Thus, $\tau$ is stipulated at the beginning of each run and does not change.
We also define the radially and temporally varying optical depths,
by subdividing the radial domain into thin strips of radial length $L_S$:
\begin{equation}
\tau(x_i, t) = N_i(t) \pi a^2/(L_S L_y), 
\end{equation}
where $x_i$ is the radial location of, and $N_i (t)$ is the number of particles in, the $i$'th strip at time $t$. 

The filling factor is defined as the proportion of volume taken up by the particles. For spherical particles it can be defined as $FF = (4\pi/3) n a^3$, where $n$ is volumetric number density. Particularly useful is the filling factor at the mid-plane $FF_0$, which requires the calculation of the number density at $z = 0$.

The mean velocity dispersion tensor is computed via
\begin{equation}
W_{ij} = \langle \dot{x}_i \dot{x}_j \rangle
\end{equation}
where $(\dot{x}_1,\dot{x}_2,\dot{x}_3)=(\dot{x},\dot{y}+\tfrac{3}{2}\Omega x,\dot{z})$ is the velocity relative to the 
shear and the angle brackets indicate a suitable average over the particles and possibly over time. The velocity dispersion $c^2$ is then $W_{ii}/3$. 
Note that this definition is only correct if there are no mean flows additional to the Keplerian shear. If such flows are slow (as in viscous instability), the error will be small, however.

The translational (local) component of the kinematic viscosity is
\begin{align}
\nu_\text{trans} = (2/3)W_{xy}/\Omega. 
\end{align}
The collisional (non-local) component of the viscosity is
\begin{align}
\nu_\text{coll} = \frac{2}{3\Omega N \delta t}\sum (x_\rangle -x_\langle )\Delta p_y
\end{align}
where the sum is taken over all binary collisions that occur in a time interval $\delta t$. Here $M$ is the total mass of all ring particles, $\Delta p_y$ is the transfer of specific $y$ momentum from the inner to the outer particle in each collision, and $x_\rangle$ and $x_\langle$ are the radial locations of the two impacting particles (Wisdom \& Tremaine 1988; Daisaka, Tanaka \& Ida 2001). 
As we neglect self-gravity, there is no gravitational or wake contribution to the overall momentum transport. The total viscosity is hence $\nu_\text{tot}=\nu_\text{trans}+\nu_\text{coll}$.

To determine the thermal conductivity of a given equilibrium state, we follow the method of Salo et al. (2001) and create a steady non-uniform temperature $T$ profile in the radial ($x$) direction, where $T=c^2$. 
In our cold-state simulations, we achieve this by making $v_\text{crit}$  radially dependent in the collision law.  
In our hot-state simulations, we vary $\epsilon_\text{max}$ by a small amount in the radial direction. 
In either case, we end up with a steady-state sinusoidal radial temperature profile, though 
some experimentation is required to find the right amplitude for the variations in $v_\text{crit}$ and $\epsilon_\text{max}$.
The goal is to keep the perturbations in the temperature $\Delta T$ small, but not too small so that they are dominated by shot noise.
We typically use a simulation with $L_x=L_y=200a$ and run it for at least 1000 orbits.

After setting up the nonuniform temperature profiles, we then measure specific translational (local) and collisional (non-local) heat fluxes,
\begin{align}
    q_i^{\rm trans} &= \frac12 \sigma \langle c^2 c_i \rangle\\
    q_i^{\rm coll} &= \frac {\sigma \sum \Delta x_i \delta E^s}{N\delta t}
\end{align}
where $\sigma=N/(L_x L_y)$ is the number surface density, $\delta x_i$ is the absolute difference of the $i$-coordinates of the two particles involved in a collisions, and $\delta E^s$ is the change in transported energy (as opposed to dissipated energy) during the collision for the particle with the larger $x_i$ coordinate. Finally, we assume the heat flux is linearly dependent on the temperature gradient,
\begin{align}
    {\mathbf q} = -\kappa {\mathbf \nabla} T.
\end{align}
We can then correlate the measured $q_x$ and $\d_x T$ and retrieve the conductivity $\kappa$ using a least squares fit. Finally, to verify our set up was working properly, we successfully reproduced Fig.~8 in Salo et al.~(2001), though omit these results for the sake of space.

\subsubsection{Parameters and initial conditions}

In all our $N$-body simulations, we adopt units so that $a=1$ and $\Omega=1$, though in what follows $a$ and $\Omega$ reappear occasionally in order to make a point.
As a consequence, the main physically relevant input is the collision law. 
Specifically, we have some combination of $v_\text{crit}/(a\Omega)$, $b$, and $\epsilon_\text{max}$ for non-constant collision laws. We also have the sizes of the numerical domain $L_x$ and $L_y$ and a constant dimensionless 
time-step $\Omega dt$.

We use initial conditions where particles are arranged uniformly in the plane with a uniform optical depth $\tau$. 
Therefore an important initial input is particle number $N$ while keeping the computational domain fixed. Particles are normally distributed in the $z$-direction. The initial velocities are also normally distributed with an initial velocity dispersion $c_0$. 
In most cases we initialize the particles close to the thermal equilibrium we believe to be present. 

We present convergence tests in Appendix \ref{sec:convergencetests}. 
These tests shows that our simulations are converged as we vary numerical parameters for both extremely
high and low optical depth, as well as hot and cold equilibria. 
For the regimes that we are interested in, we found that a dimensionless timestep of $10^{-3}$ and a box size of 10s to 100s particle radii are sufficient. 
The large box sizes are needed only for very hot and dilute rings. 

\begin{figure}
    \centering
    \includegraphics[width=\columnwidth]{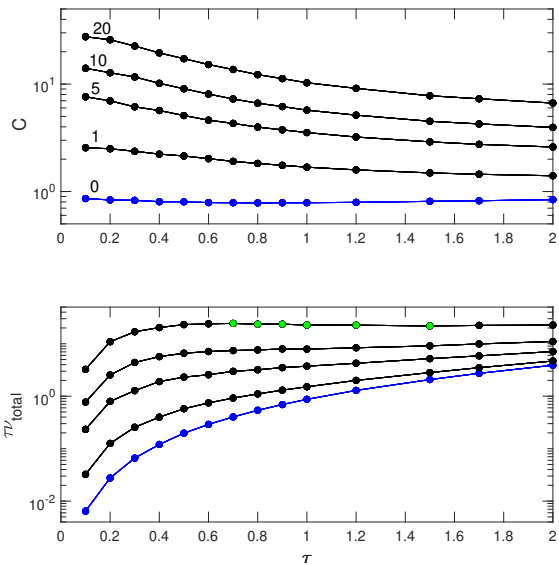}
    \caption{Velocity dispersion and total angular momentum flux $\tau\nu_\text{tot}$ versus optical depth $\tau$ for various hard-sphere $\epsilon$ laws, calculated from $N$-body simulations. In the top panel the appended numbers `1-20' describe the values of $v_\text{crit}/(a\Omega)$ when using the standard Bridges law, whereas `0' indicates runs with a constant $\epsilon=0$. In the bottom panel, the ordering of the curves is retained. The green symbols indicate that the viscous flux is decreasing and the disk viscously unstable.} 
    \label{previouswork}
\end{figure}

\subsection{Kinetic theory}

Though not the focus of this paper, it is useful to have some kinetic theoretical results, especially as they reveal the existence of the additional (thermally unstable) middle branch of equilibrium solutions.
The formalism adopted is Latter and Ogilvie's (2008) reformulation of Araki and Tremaine (1986), which does not attempt to solve the Boltzmann-Enskog equation but rather a truncated moment hierarchy of continuum equations. 

In previous deployments of this approach, the dependence of $\epsilon$ on the impact speed was only approximately incorporated via a `pre-averaging' procedure (see Section 2.2.7 in Latter and Ogilvie 2008). Though convenient, this introduces unacceptable errors when using complicated non-monotonic laws as in Section 2.1.3. Thus the complete formalism is adopted. This does require completing three (instead of two) integrations in the collision term. The other main approximations adopted are `vertical locality' and a triaxial Gaussian for the velocity ellipsoid (see Araki and Tremaine 1986 and Latter and Ogilvie 2008 for more details). 

\section{Homogeneous steady states}

In this section we simulate various thermodynamic equilibria and demonstrate that a non-monotonic epsilon law supports up to two equilibria for a given optical depth. We characterise these several states with respect to not only their velocity dispersion, but also their packing fraction $FF_0$ and transport properties, especially with respect to angular momentum and heat.

We begin by reproducing previous results in the literature with both a constant and monotonic epsilon law so as to verify that our code is working properly. Moreover, as argued in Section \ref{bistability}, some of the equilibria obtained are limiting cases of those appearing in the bistable circumstances explored later and are thus useful in setting the scene.

\begin{figure*}
    \includegraphics[width=\textwidth]{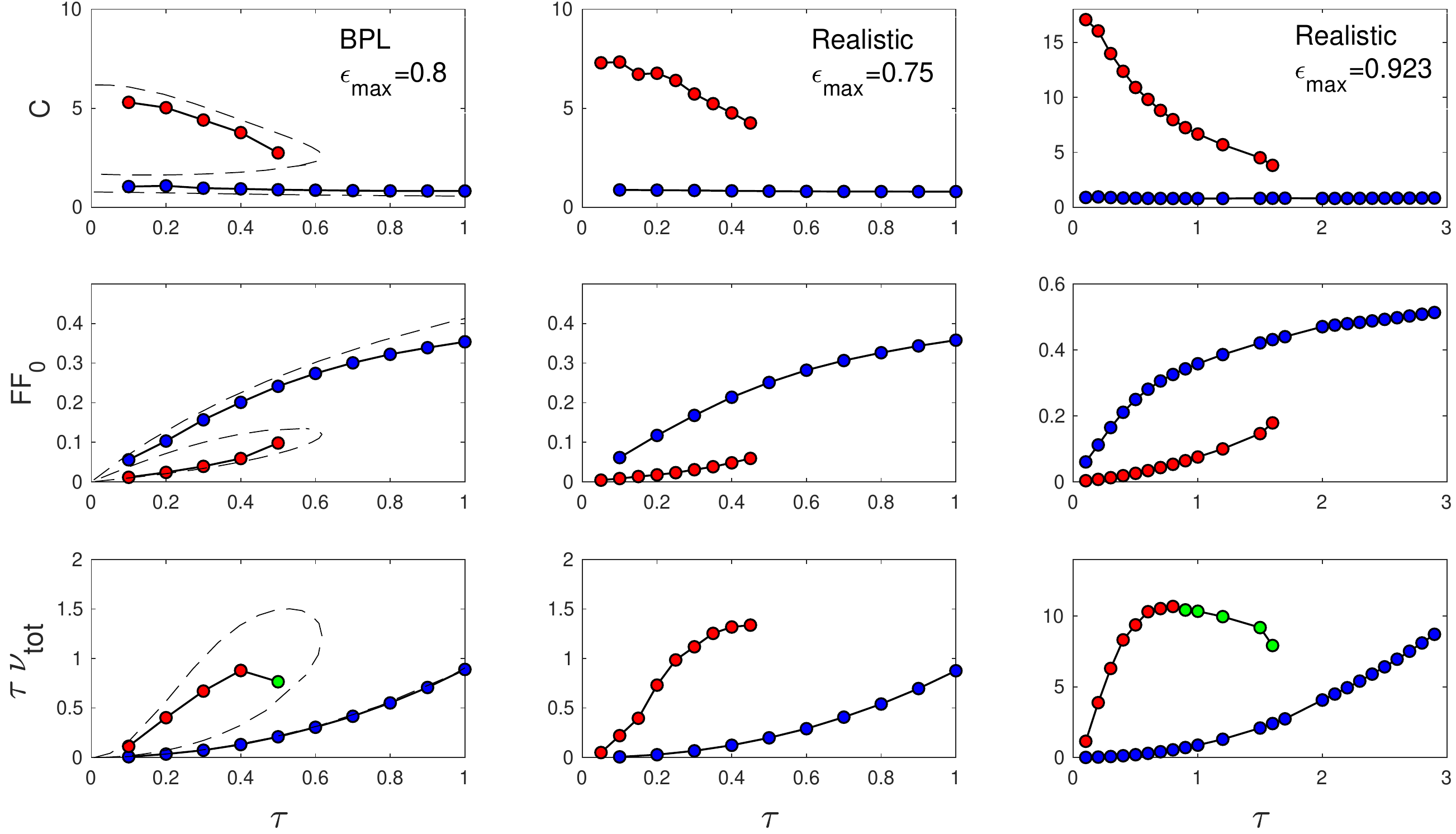}
    \caption{Selected equilibrium properties as functions of $\tau$ for three regolith $\epsilon$-laws (the three columns). The leftmost column shows equilibria computed with the broken-power law model (BPL) with $\epsilon_0=0$ and $\epsilon_\text{max}=0.8$, whereas the other two columns show the realistic model with $\epsilon_\text{max}=0.75$ and $0.923$. In all cases $v_\text{crit}=5$.
    In the top row the joined circles denote the velocity dispersion calculated by $N$-body simulations, with the colours indicating hot (red) or cold (blue) branches.  The second and third rows show the filling factor and total angular momentum flux respectively. The dashed curve indicates equivalent solutions obtained from the kinetic theory (in the BPL case only). In the bottom row, a green symbol indicates expected viscous instability.}
    \label{hysteresis}
\end{figure*}

\begin{figure}
    \centering
    \includegraphics[width=0.45\textwidth]{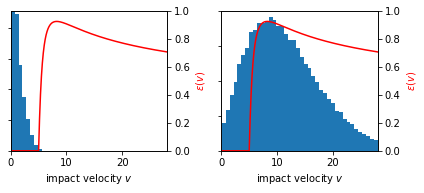}
    \caption{The distribution of impact velocities in simulations using the `realistic' law at $\tau=1$ with parameters $v_\text{crit}=5$, $b=1$, and $\epsilon_\text{max}=0.923$. The left panel shows the system in the cold state. The right panel shows the system in the hot state. The red line corresponds to the $\epsilon$ law adopted.} \label{distribution}
\end{figure}

\subsection{Comparison with previous calculations}

Our reference cases include the simulations of Salo (1991), who employed a Bridges law but with a variable scale velocity, i.e. Eq.\eqref{BPL} with $\epsilon_0=1$ and $v_\text{crit}=1,5,10,20$ (see Section 2.1), and also simulations with a constant $\epsilon=0$, which brings about a very cold state. The results of our calculations are plotted in Fig.~\ref{previouswork}, in which we show the velocity dispersion $c$ and angular momentum flux $\tau \nu_\text{tot}$ versus optical depth $\tau$. The simulations were run until they were collisionally relaxed, and then continued for the same length of time to obtain averaged quantities. When $\tau$ was low, and collisions relatively infrequent, the total run time was $>1000$ $\Omega^{-1}$; but at higher $\tau$ ($\sim 2$) runs could be as a short as 50-80 $\Omega^{-1}$. 

Direct comparison of Fig.~\ref{previouswork} with the numerical results of Salo (1991; cf. his Figs 3-5) shows good agreement, and also consistency with the kinetic theory of Latter \& Ogilvie (2008) (note that both these works denote $v_\text{crit}$ by $v_b$). An interesting feature of the `warmer' solution branches is the decreasing viscosity with $\tau$. In fact, the hottest case, $v_\text{crit}=20$, is viscously unstable because the gradient of the angular momentum flux $\tau\nu$ is negative in an interval of $\tau$ (green markers).

By inflating $v_\text{crit}$ in the Bridges law the velocity dispersion of the system can be controlled and, in particular, set to `warm' values greater than $a\Omega$ and, consequently, greater than the temperature of the very cold $\epsilon=0$ states. These warm and cold states help illustrate the arguments presented in Section \ref{bistability}. If we take one of the two non-monotonic collision laws and set $v_\text{crit}$ ten or more times $a\Omega$, then start the simulation with a hot initial condition, we might expect the subsequent spread of impact speeds to be sufficiently far from $\epsilon$'s turning point (cf. Fig.~\ref{epsilons}) so that the system settles into a warm `Bridges equilibrium', similar to those plotted in Fig.~\ref{previouswork}. On the other hand, if we begin the same simulation but with very cold initial velocities ($\ll v_\text{crit}$), the subsequent spread of impact speeds will remain less than $v_\text{crit}$ and $\epsilon$ will almost always take the value of 0; the system will then converge to the appropriate constant $\epsilon=0$ state in Fig.~\ref{previouswork}. 
  Note that the Bridges law produces a velocity dispersion $c$ that decreases with $\tau$ and we may then expect that for sufficiently large $\tau$ the upper `hot state' will be too close to the `cold state' and bistability may disappear.

\subsection{Non-monotonic collision laws}

In this section we calculate equilibria for `regolith' epsilon laws that are non-monotonic: either the broken power law (BPL) with $\epsilon_0=0$ or the realistic law \eqref{regolith}. The parameters are $\epsilon_\text{max}=0.75, 0.8$ or $0.923$, $v_\text{crit}=5a\Omega$, and $b=1$, though we examine a broader spread of values in Section \ref{survey}. We first examine in some detail the thermal properties of the states, then their transport of angular momentum and heat.

\subsubsection{Thermal hysteresis}

Figure 3 constitutes the first main results of the paper. Here we plot the equilibrium velocity dispersions (top row), filling factors (middle row), and total radial angular momentum fluxes ($\tau\nu_\text{tot}$; bottom row) obtained in a sequence of simulations at different optical depths and for different $\epsilon$ models and parameters. Each circular marker corresponds to a different simulation. These values are obtained by time averaging a quantity once the system has become collisionally mature, as earlier. For example, $\tau=0.1$ runs were run for 1600$\Omega^{-1}$ and averaged for the last 800$\Omega^{-1}$, while at $\tau=2$ the total run length was 80$\Omega^{-1}$, with the averaging taking place over the last 40$\Omega^{-1}$.

As is clear, in the three models presented, two steady state branches (distinguished by red and blue) are possible within a certain range of optical depth. Which of the two the system selects depends on the initial condition: a `cold start' (low initial $c$) usually (but not always) takes the system to the nearby cold state, whereas a `hot start' (initial $c$ sufficiently high) settles on the hot state. Typically, runs starting with $c= 0.5a\Omega$ converged to the nearby cold state, while runs beginning with $c=10a\Omega$ migrated to the hot state, if one was available, even if that state's velocity dispersion was significantly larger than the initial $c$. The direction of migration is discussed further in Section 4.

The apparent bistability extends over a range of small to intermediate optical depths. Beyond a special $\tau$ the hot state disappears, and all hot start simulations  landed on the cold branch. At small $\tau$ we never found that the cold state disappeared, except in the case of the realistic model with $\epsilon_\text{max}=0.923$ and $\tau=0.1$; this equilibrium was metastable (explored in more detail in Section 4). The bistable regime's width (in $\tau$) depends on the parameters. From Fig. 3, increasing the $\epsilon_\text{max}$ in the realistic model from 0.75 to 0.923 moved the special $\tau$ from roughly 0.5 to 1.6 (cf.\ middle and right columns).

\begin{figure}
    \centering
    \includegraphics[width=0.5\textwidth]{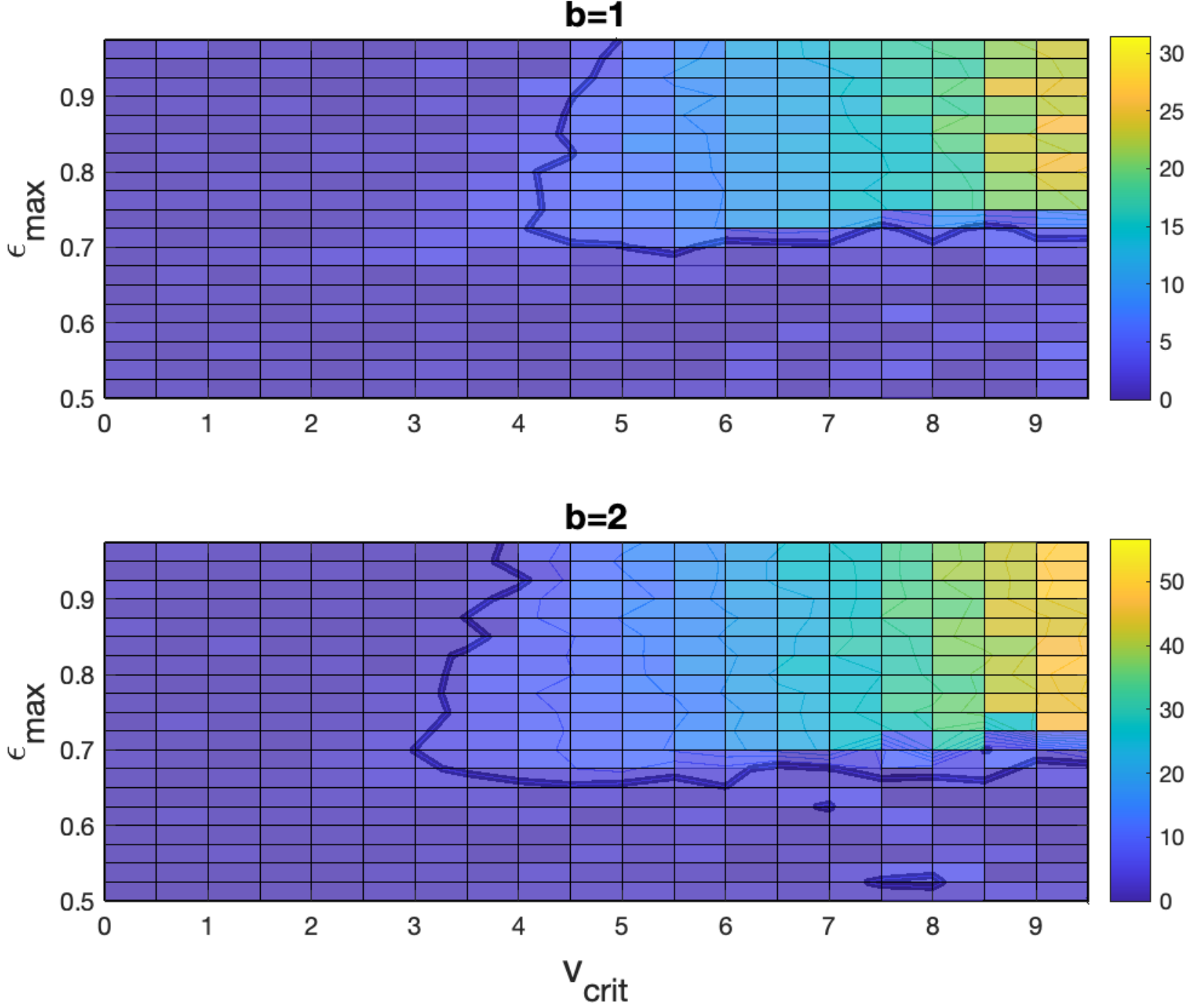}
    \caption{Grids of simulations undertaken with different $v_\text{crit}$ and $\epsilon_\text{max}$ using the realistic regolith law with widths $b=1$ (top) and $b=2$ (bottom). Colours correspond to values of $|c_\text{hot}-c_\text{cold}|$ (see text). The contour is a conservative boundary between cases that support bistability (to the right and above) and those that do not.}
    \label{bigrid}
\end{figure}

The cold equilibria take $c$ values very much in agreement with the constant $\epsilon=0$ states simulated in the previous subsection, while the hot state resembles a Bridges law, with $c$ decreasing with $\tau$. In fact, the hot simulations of the realistic model with $\epsilon_\text{max}=0.923$ take a similar $c$ as the Bridges $v_\text{crit}=10$ runs, while those  with $\epsilon_\text{max}=0.75$ resemble a Bridges law with (roughly) $v_\text{crit}=5$. These similarities bolster our interpretation of the two states as `separated' by the turning point of the $\epsilon$ curve: only a minority of collisions in the hot state occur with the low impact speeds that would trigger $\epsilon=0$, while collisions in the cold state rarely occur with impact speeds sufficiently large to trigger larger $\epsilon$. To flesh out this point further we plot in Fig.~\ref{distribution} the distribution function of impact speed for a hot state (right panel) and a cold state (left panel) for the same $\tau=1$ (and other parameters). Superimposed in red is the $\epsilon$ law used. As the left panel indicates, cold state collisions are almost completely inelastic; the narrow spread in impact speeds barely overlaps the portion of the curve for which $\epsilon\neq 0$.
In contrast, the hot state (shown in the right panel) is much broader and thus samples a wide range of $\epsilon$, but importantly peaks at speeds which yield collisions with a small dissipation of energy.

The filling factors in the middle row of Fig.~3 reveal that the hot branches are far less dense than the cold branches. For example, in the realistic model with $\epsilon=0.923$, at $\tau=1$ the hot state possesses a filling factor of 0.08, while the cold state has 0.35. The difference, of course, is not due to the surface number density (which is the same) but because the disk semi-thickness is so different between these two states: in the hot state it is $\approx 6a$, compared to $\sim a$ in the cold state. The ratio of the two filling factors should scale roughly with the ratio of semi-thicknesses and that is indeed what we see.

The hot state branch terminates when its velocity dispersion approaches a critical value $\sim 3$. In reality the system here encounters a saddle-node bifurcation and the solution curve bends `backwards' thus forming an intermediate branch of thermally unstable solutions. Because these solutions are unstable they cannot manifest in $N$-body simulations\footnote{See Salo et al.~(1988) for a numerical exploration of a thermally unstable state.}, but they can be calculated by kinetic theory. Kinetic theoretical equilibria are plotted in the leftmost column with a dashed black curve; the top and middle panels show clearly an intermediate cool, semi-dense branch. The agreement between theory and simulations is qualitative good, with the biggest deviation in the translational viscosity in the hot state, a discrepancy that has been noted in previous comparisons (Latter and Ogilvie 2008, Rein and Latter 2013).\footnote{Unfortunately, numerical difficulties prevented us calculating kinetic solutions for the realistic model.}

\begin{figure*}
    \centering
    \includegraphics[width=0.99\textwidth]{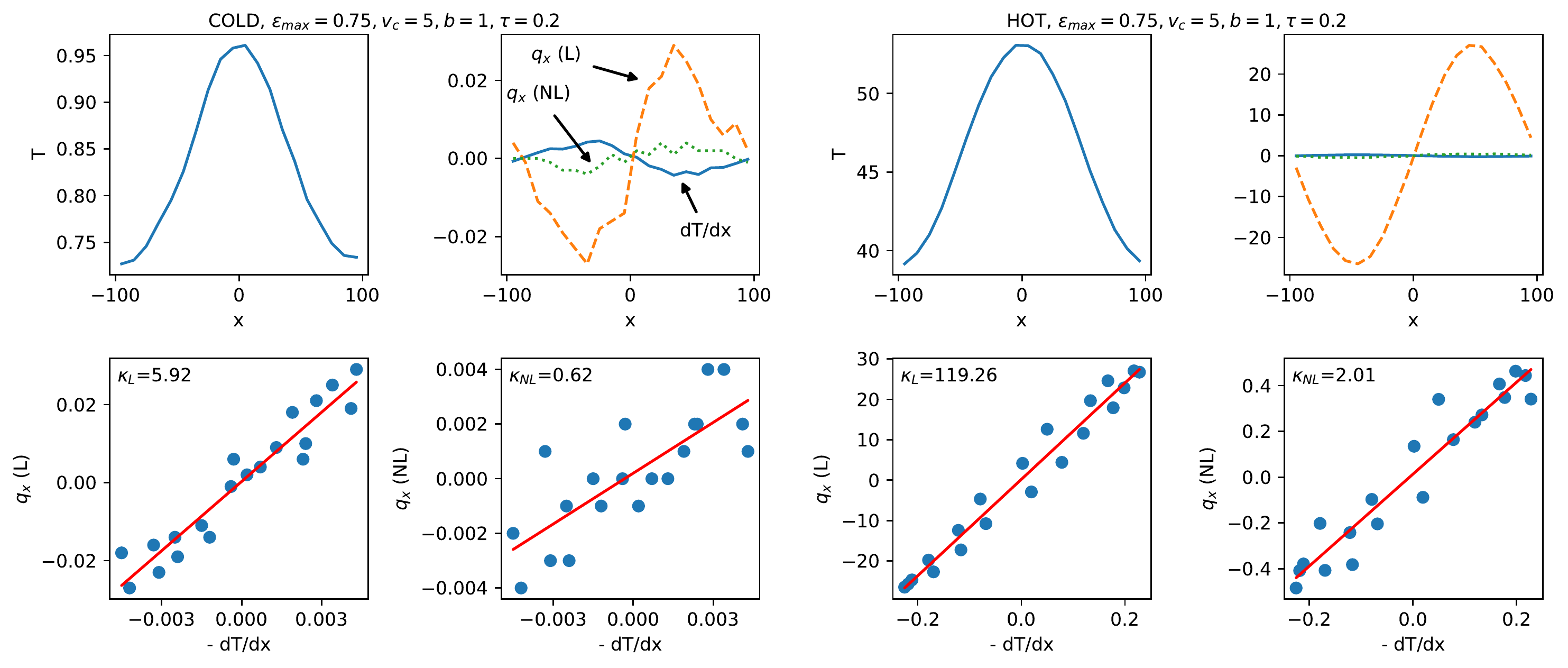}
    \caption{Thermal diffusivity measurements for $\tau=0.2$ in the cold state (left panels) and hot state (right panels) for the realistic model with $\epsilon_\text{max}=0.75$, $v_\text{crit}=5$, and $b=1$. 
    \label{fig:diffusivity}
    }
\end{figure*}

\subsubsection{Parameter survey}
\label{survey}

In the preceding subsection we examined only three parameter sets/models; in this subsection we adopt the realistic $\epsilon$ law and scan through $v_\text{crit}$ and $\epsilon_\text{max}$ for two different widths $b$. Our aim is to determine how representative the thermal hysteresis explored in the previous subsection really is.
Of particular interest are the lowest values of $v_\text{crit}$ and $\epsilon_\text{max}$ that yield bistability. 

In Fig.~\ref{bigrid} we present `bistability plots' for $b=1$ and 2. Each square in the grid corresponds to a parameter pair $(v_\text{crit},\epsilon_\text{max})$, and for each square we conduct two simulations with $\tau=0.1$, one with a hot initial condition and the other with a cold initial condition. Each simulation has been run until thermal equilibrium has been obtained, and the difference in final velocity dispersion calculated, $|c_\text{hot}-c_\text{cold}|$. Finally, the square is coloured accordingly (cf.\ the colour bar). If the difference in final $c$ is between 0 and 5, we interpret that the two simulations are converging on to the same (cold) equilibrium. Values larger than 5 (admittedly, a rather large value, given Fig.~3) we assume correspond to a bistable situation: the two simulations are settling on different thermal states. In both panels we have superimposed the contour of $|c_\text{hot}-c_\text{cold}|=5$. The reader should then assign bistability to regions of the parameter plane above and/or to the right of this curve.

The plots indicate, as expected, that bistability is favoured by larger values of $v_\text{crit}$ and $\epsilon_\text{max}$. Increasing both parameters helps to separate the typical impact speeds of the hot state from those of the cold state. Interestingly, the bistable region is quite rectangular. Thus when $b=1$, bistability is guaranteed (roughly) if both $v_\text{crit}> 4$ and $\epsilon_\text{max}> 0.7$. We expect that these parameter restrictions should hold roughly for other non-monotonic laws. 
Finally, the range of bistability is also sensitive to the width of the epsilon law, as the $b=2$ plot demonstrates. Increasing the width also helps separate out the two states. In the $b=2$ case bistability occurs when $v_\text{crit}>3$ and $\epsilon_\text{max}>0.65$. 

\subsubsection{Viscous properties}

The equilibrium states discussed in the previous subsection support a viscous stress that, by acting on the background orbital shear, transports angular momentum radially across the numerical domain. The viscous properties of the flow are important thermodynamically because the stress extracts free energy from the shear, thus providing the heating source in the thermal balances undergirding these states. But the viscous stress is also important dynamically because it can beget instabilities, such as the viscous overstability and instability (Schmidt et al.~2009). In particular, if $d(\tau\nu_\text{tot})/d\tau$ is negative then viscous instability occurs (Lin and Bodenheimer 1981, Lukkari 1981, Ward 1981). 




The angular momentum flux is plotted in the bottom row of Fig.~3. Note that a subset of hot states possess a decreasing flux and are thus viscously unstable; these are marked in green. In the BPL model, the unstable interval encompasses $\tau$ of 0.4 and 0.5, whereas in the realistic model only the $\epsilon_\text{max}=0.923$ case yields instability and then for $\tau$ between approximately 0.8 and 1.6. Instability here is associated with a dominant \emph{translational} viscosity, which can decline at sufficiently large $\tau$. Growing modes do not appear in these simulations, however, because the numerical domain size is smaller than the shortest unstable wavelength; in Section 5.2.2 we simulate larger domains and recover the instability.

\begin{table}
\begin{tabular}{|l || l | l || l | l |}
 \hline
 $\tau$ & $\kappa_\text{L}$ (C) & $\kappa_\text{NL}$ (C)  & $\kappa_\text{L}$ (H) & $\kappa_\text{NL}$ (H)  \\
 \hline
\hline
0.1 & 4.75 & 0.42 & 77.94 & 0.72 \\
0.2 & 5.92 & 0.62 & 119.26 & 2.01 \\
0.3 & 7.42 & 1.15 & 111.88 & 3.39 \\
0.4 & 6.83 & 1.61 & 89.95 & 3.59\\
\hline
\end{tabular}
\caption{Calculated translational (local) thermal conductivities $\kappa_\text{L}$ and collisional (non-local) thermal conductivities $\kappa_\text{NL}$ in cold (C) and hot (H) equilibria at various optical depths $\tau$. A realistic collision law is adopted with $\epsilon_\text{max}=0.75$, $v_\text{crit}=5$, and $b=1$.}
\end{table}

\subsubsection{Thermal conductivity}

Anticipating later sections which explore different thermal states that spatially adjoin, we compute the radial flux of thermal energy. In the absence of any mean spatial gradients, such as in the homogeneous equilibria calculated, the flux must be zero. But if two states connect in radius the flux must control, in part, how their interface evolves. 

As explained in Section 2.3.3, we adopt the approach of Salo et al~(2001) and impose a radial sinusoidal temperature structure upon the box, through the parameters $v_\text{crit}$ and $\epsilon_\text{max}$. In Fig. 7 we show calculations of the radial thermal flux $q_x$ and the thermal conductivity $\kappa$ for a fixed set of parameters ($\epsilon_\text{max}=0.75$, $v_\text{crit}=5$, $b=1$) and for the same optical depth $\tau=0.2$. The left four panels correspond to the cold state ($c\approx 1$), and the right to the hot state ($c\approx 6$). The top left panel in each case describes the temperature profile across the box, while the top right panel shows the temperature gradient (solid blue), the translational (local, `L') heat flux (dashed gold), and the collisional (nonlocal, `NL') heat flux (dotted green). The latter two are plotted separately as functions of the temperature gradient in the bottom panels; a best-fit line extracts the conductivities.  

In both the hot and cold cases, the translational heat flux dominates the collisional flux. This means that the heat flux in the two states differs significantly, despite possessing the same $\tau$.  In Table I we list $\kappa$ for a range of $\tau$ and otherwise with the same parameters as in Fig.~6. 


\begin{figure*}
    \centering
    \includegraphics[width=0.75\textwidth]{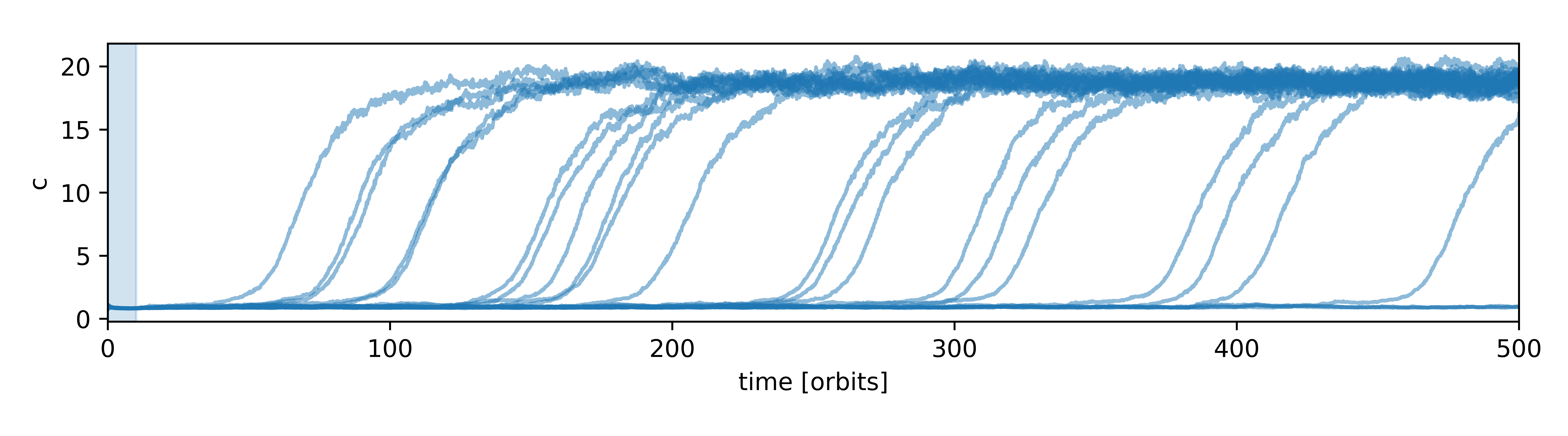}
    \includegraphics[width=0.75\textwidth]{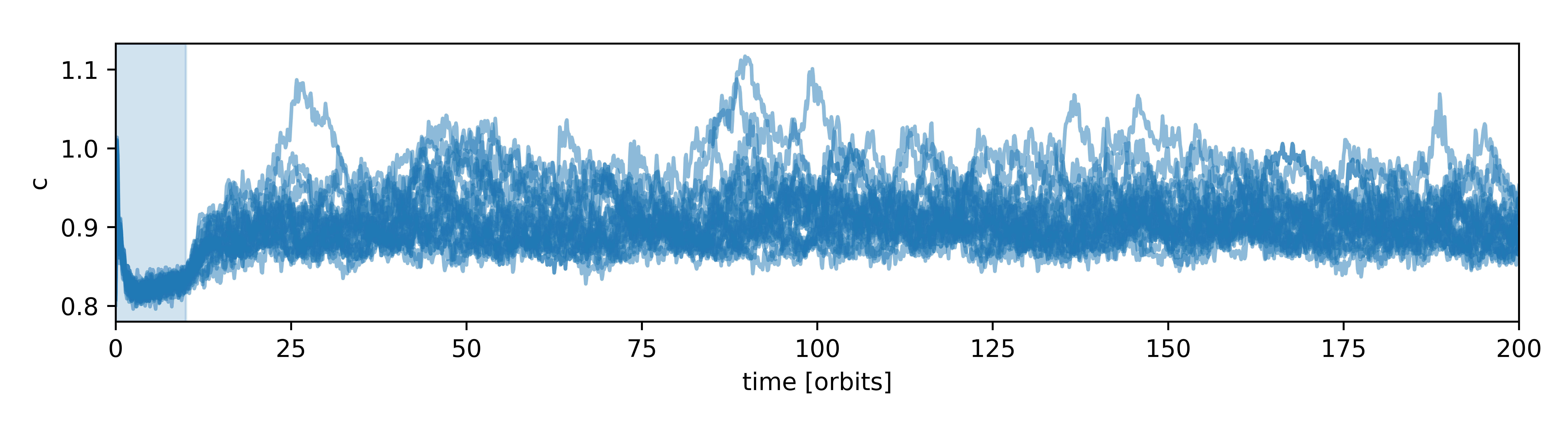}
    \caption{Velocity dispersion as a function of time for runs with $\tau=0.1$ (top panel) and $\tau=0.2$ (bottom panel). The realistic model is adopted with $\epsilon_\text{max}=0.923$, $v_\text{crit}=5$, and $b=1$.
    \label{fig:trans}
    }
\end{figure*}

\begin{figure*}
    \centering
    \includegraphics[width=0.75\textwidth]{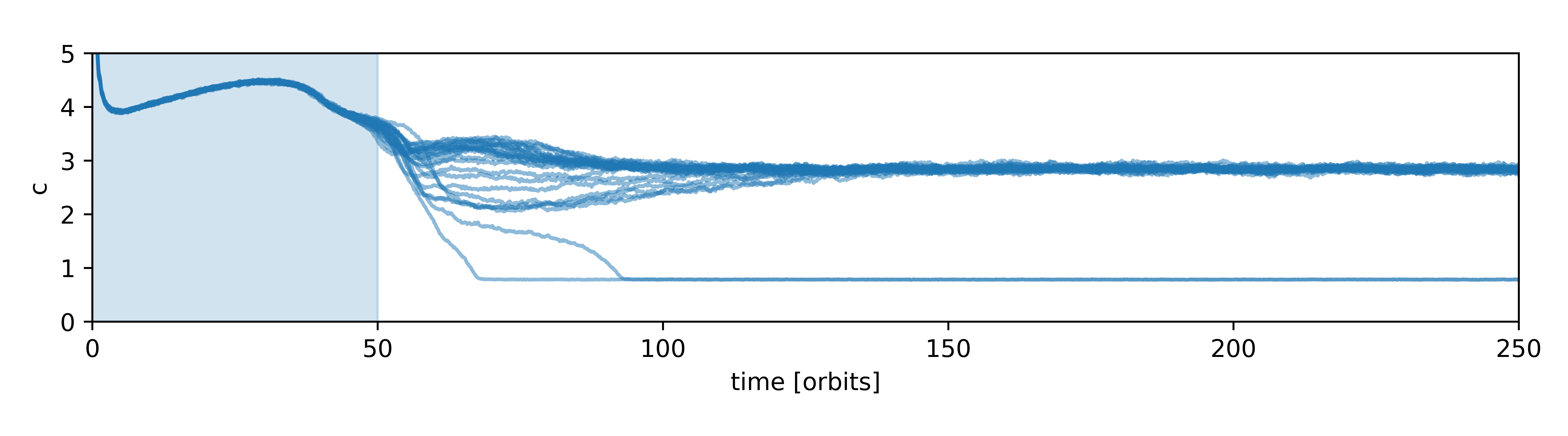}
    \includegraphics[width=0.75\textwidth]{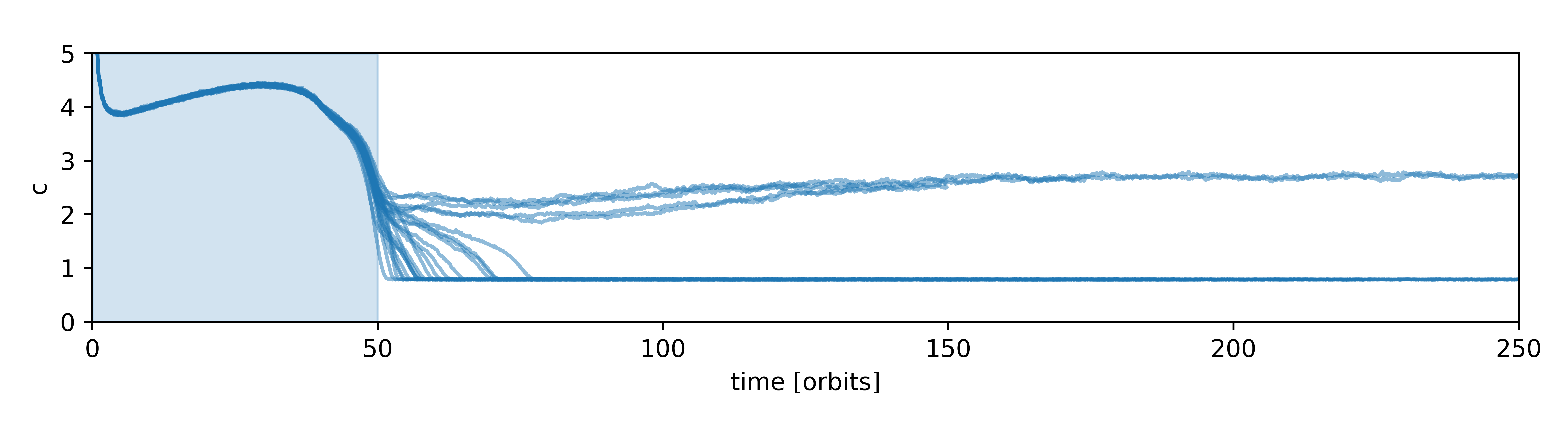}
    \caption{Velocity dispersion as a function of time for runs of different initial conditions with $\tau=1.61$ (top panel) and $\tau=1.64$ (bottom panel). The realistic model is adopted with $\epsilon_\text{max}=0.923$, $v_\text{crit}=5$, and $b=1$.
    \label{fig:trans3}
    }
\end{figure*}


\section{Metastability}

In the last section we calculated steady states that appear to be thermally stable, at least linearly according to a continuum interpretation. However, $N$-body systems are replete with small but finite amplitude shot noise that continually tests the \emph{nonlinear} stability of any steady state. If the basin of attraction of a linearly stable state is small relative to the amplitude of these fluctuations, the system can potentially jump out of the state and migrate elsewhere. Many physical and biological systems offer similar examples of noise destabilising what should be linearly stable fixed points (e.g. Mel'nikov 1991, May 1973, De Swart and Grasman 1987, Majda, Timofeyev and Vanden-Eijinden 1999, 2003). In this section we investigate this possibility.

Our focus will be on cold states of low-optical depth and on the hot states near the saddle node bifurcation. The reason is that these states are close to the unstable middle branch which can serve as the boundary of the basin of attraction in each case. We find that, for the parameters and models we employ, \emph{metastability} is relatively uncommon, only occurring in certain dilute and cold states. In particular, states near the saddle node are generally stable to shot noise perturbations. 

Before presenting our results we emphasise that we only explore the effect of intrinsic shot noise, but in real rings there are several other sources of finite amplitude disturbances that may work similarly, e.g. meteoroid bombardment, embedded moonlets, density waves, and gravity wakes.

\subsection{Cold to hot transitions}

We find spontaneous transitions from the cold lower branch to the hot upper branch in only a few low $\tau$ cases when adopting a realistic collision law and $\epsilon_\text{max}=0.923$. Specifically, when 
$\tau=0.1$ the system can hover about the cold steady state for several hundred orbits before jumping to the hot state. 

To probe this behaviour we ran 24 runs with slightly different initial conditions (varying both particles' locations and velocities) but all starting with the same low $c$. To make doubly certain that the system is as close to the cold equilibrium as possible, and that any future transition is not the result of a wayward initial condition, we force $\epsilon=0$ (a constant) for several orbits at the start. 

The evolution of these runs are plotted in the top panel of Fig. \ref{fig:trans}, with the shaded region indicating when $\epsilon=0$.
As is clear from the figure, all but three runs jumped to the hot state by 500 orbits (roughly $>25$ collision times), though there was a wide spread of transition times, indicative that the process is stochastic and issues from the noise: ultimately, after some period, an overenthusiastic collision, dissipating insufficient velocity dispersion, seeds a patch of more energetic particles, which then spreads spatially and takes over the system. 



Of course, this is only part of the story, because energetic events must happen at slightly larger $\tau$ but do not appear to instigate runaway heating. Indeed, we undertake a similar experiment at $\tau=0.2$, plotted in the lower panel of Fig. \ref{fig:trans}, and witness no transitions at all. What is key is the overall basin of attraction of the cold state; as shown by the kinetic curves in the top left panel of Fig.~3, the middle unstable branch and the cold lower branch become closest at low $\tau$. The middle branch acts as the boundary of the lower state's basin of attraction (at least in this simple phase space projection); thus at low $\tau$ it becomes more likely that a finite amplitude perturbation can tip the system over this boundary. That said, it is not straightforward to firmly connect microphysical fluctuations (shot noise) to such a mean finite-amplitude perturbation in this phase space.

\subsection{Hot to cold transitions}

We now check if it is possible to obtain spontaneous hot to cold transitions. We focus on states near the tip of the saddle node, i.e. the termination of the hot branch (see top row in Fig. 3), and examine a range of $\tau$ between $1.61$ to $1.65$ in the realistic model with $\epsilon_\text{max}=0.923$. We simulate several runs with slightly different initial conditions, as before, and plot the results in  Fig.~\ref{fig:trans3}, top and bottom panels. 
As in the previous subsection, to ensure that we start the simulations in a hot state  we set $v_\text{crit}$ to a very small value initially. 
Over several orbits (indicated by the shaded area in the figures), we slowly increase $v_\text{crit}$ to the nominal value. 

Unlike cold to hot transitions, the systems either immediately drop to the cold state or relax into the hot state on a timescale of 10 orbits or so (a handful of collision times). At $\tau=1.65$ all the simulations ended up in the cold state, while at 1.64, some stayed in the hot state, while at lower tau again (1.61) most stay in the hot state. Putting aside the percentages in one or the other, the system transitions promptly or not at all. We attribute this more to the initial condition at the end of the blue phase, rather than having to wait for a more sluggish group of collisions that lead to a `chain reaction' and a switching of states.

The difference with the low $\tau$ runs explored earlier may partially be explained by the separation between the middle and hot branches, which is relatively large, even near the tip of the saddle node (see kinetic theory curves in top left panel of Fig.~3). Once a system settles on to the hot state, and its initial conditions mostly forgotten, its intrinsic shot noise is insufficient to tip it out of its basin of attraction and into the cold state.

\section{Thermal and viscous fronts}

Having computed several homogeneous states, we now explore the dynamics when different states spatially adjoin. If a ring region is bistable, then it is likely that such situations occur, given the varying dynamical histories at different radii. Our main focus is on the structure and evolution of the transition (or front) between two states. We will consider two cases: (a) thermal fronts, which join two states of the same $\tau$ but different $c$, and (b) viscous fronts, which connect two states of the same angular momentum flux $\tau\nu$, but different $\tau$ and $c$ 

Thermal fronts involve a hot and a cold state, with the pair joined by a vertical line in the top panels of Fig.~3. Though sharing the same optical depth, they possess distinct vertical thicknesses that may produce a photometric variation, and thus observable structure (e.g.\ Salo and Karjalainen 2003). However, the two states will support different angular momentum fluxes $\tau\nu$, and thus mass may pile up or evacuate near the thermal front, potentially leading to non-steadiness and a complete break down of the structure. We find that this is avoided if the front itself moves sufficiently fast.

One might expect radial mass redistribution is negated if two adjoining states possess the same angular momentum flux, with the pair joined by a horizontal line in the bottom panels of Fig.~3. In fact, similar structures have already been witnessed in simulations of the viscous instability with monotonic $\epsilon$ laws (Salo and Schmidt 2010). We find, however, that the finite width of the front itself spoils the exact matching of fluxes and makes the establishment of such fronts more complicated.

\begin{figure}

\centering
\includegraphics[width=0.5\textwidth]{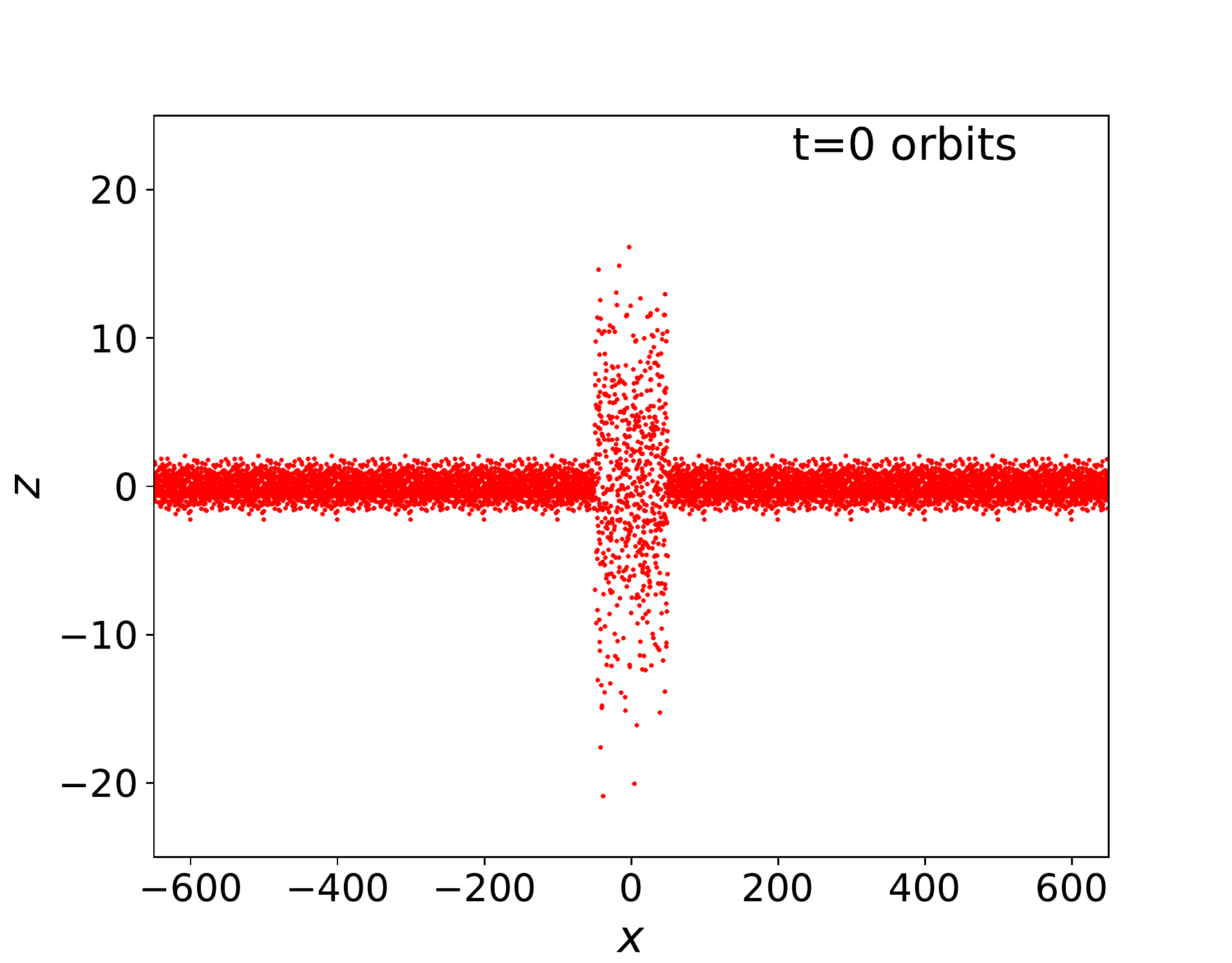}
\caption{Initial condition for the fiducial thermal-front simulation described in Section 5.1.1 in the form of an $(x,z)$ projection of the particle positions.}
\label{Ring_0}
\end{figure}

\begin{figure*}
    \centering
    \includegraphics[width=0.425\textwidth]{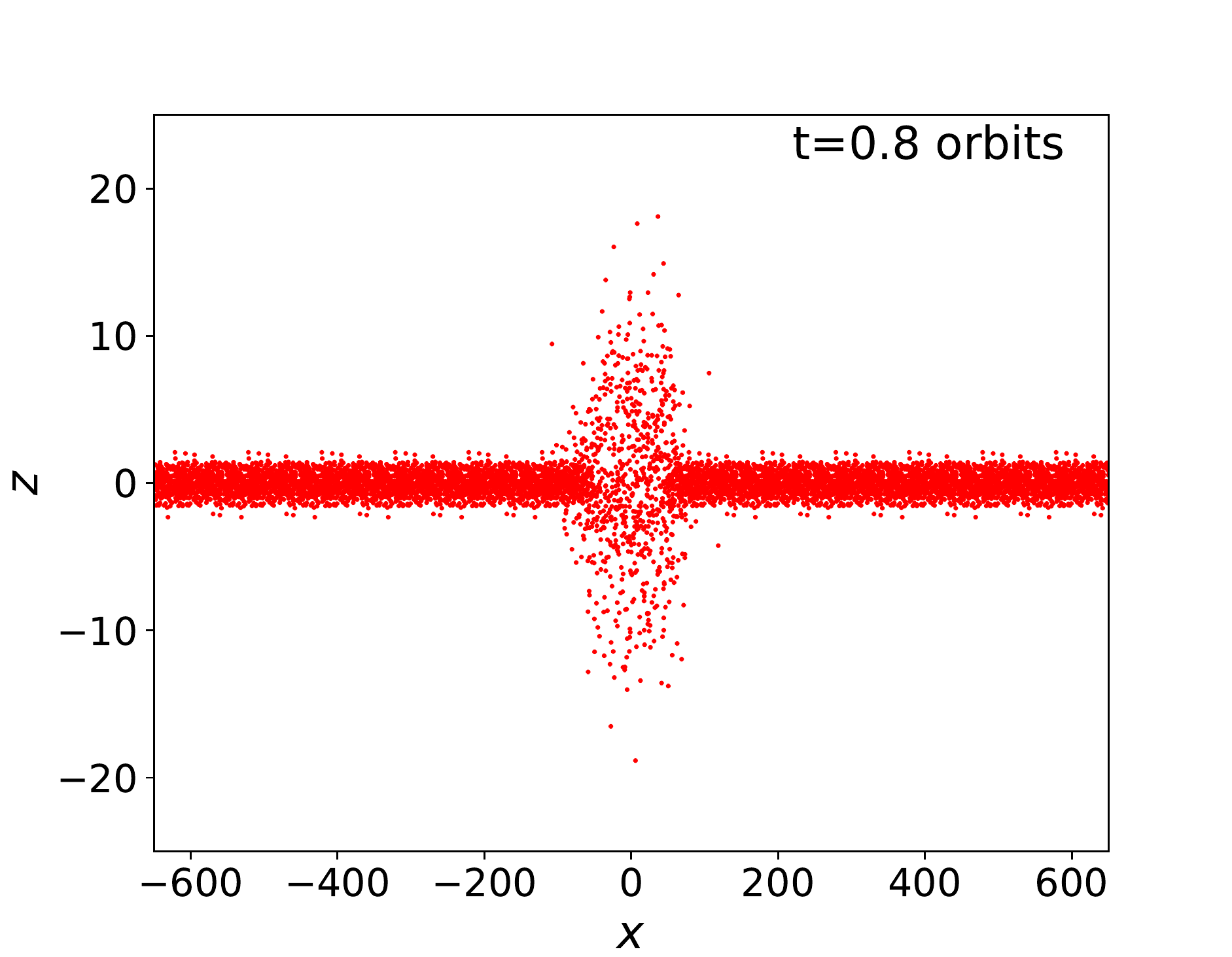}
     \includegraphics[width=0.4\textwidth]{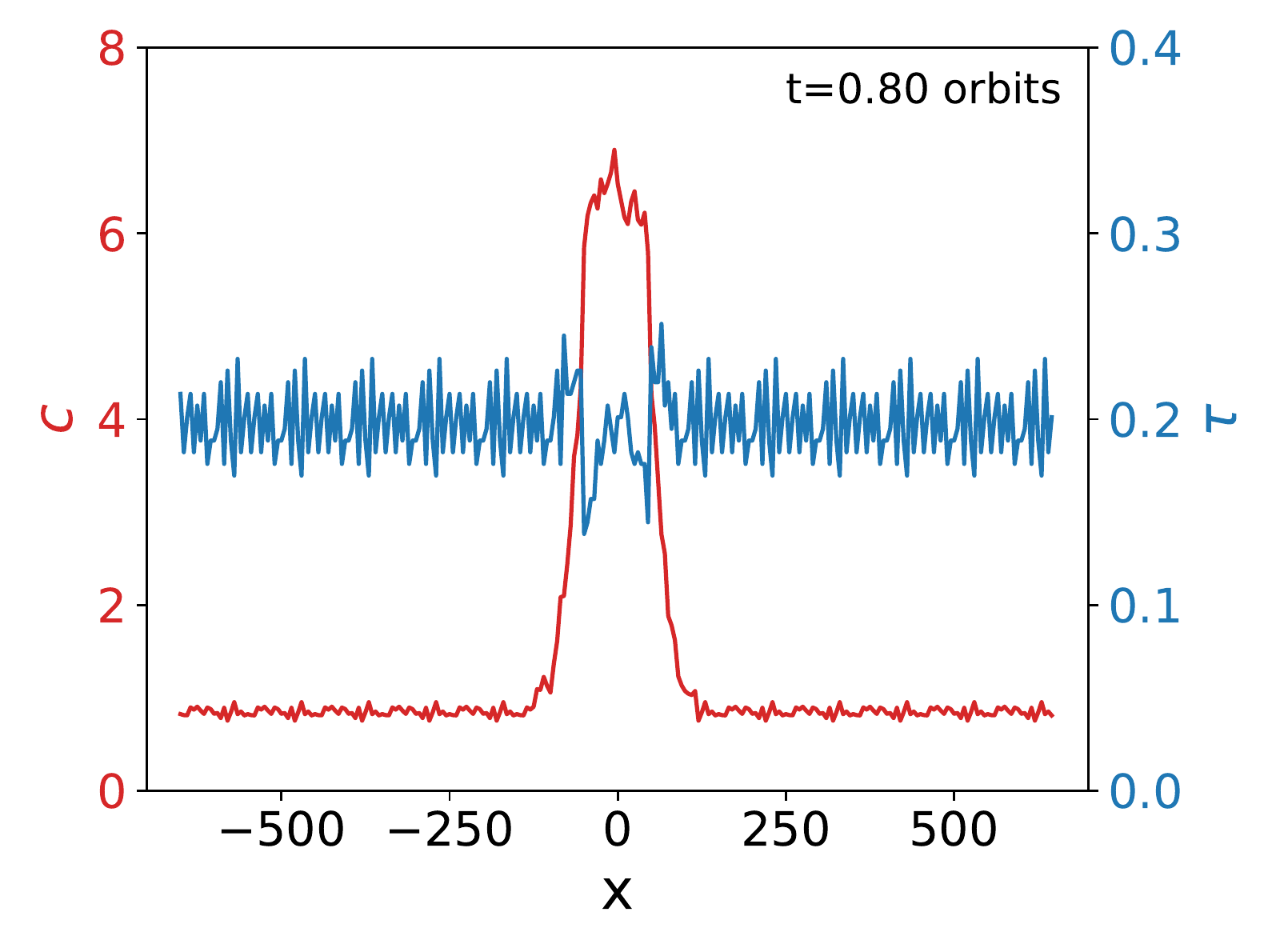}
      \includegraphics[width=0.425\textwidth]{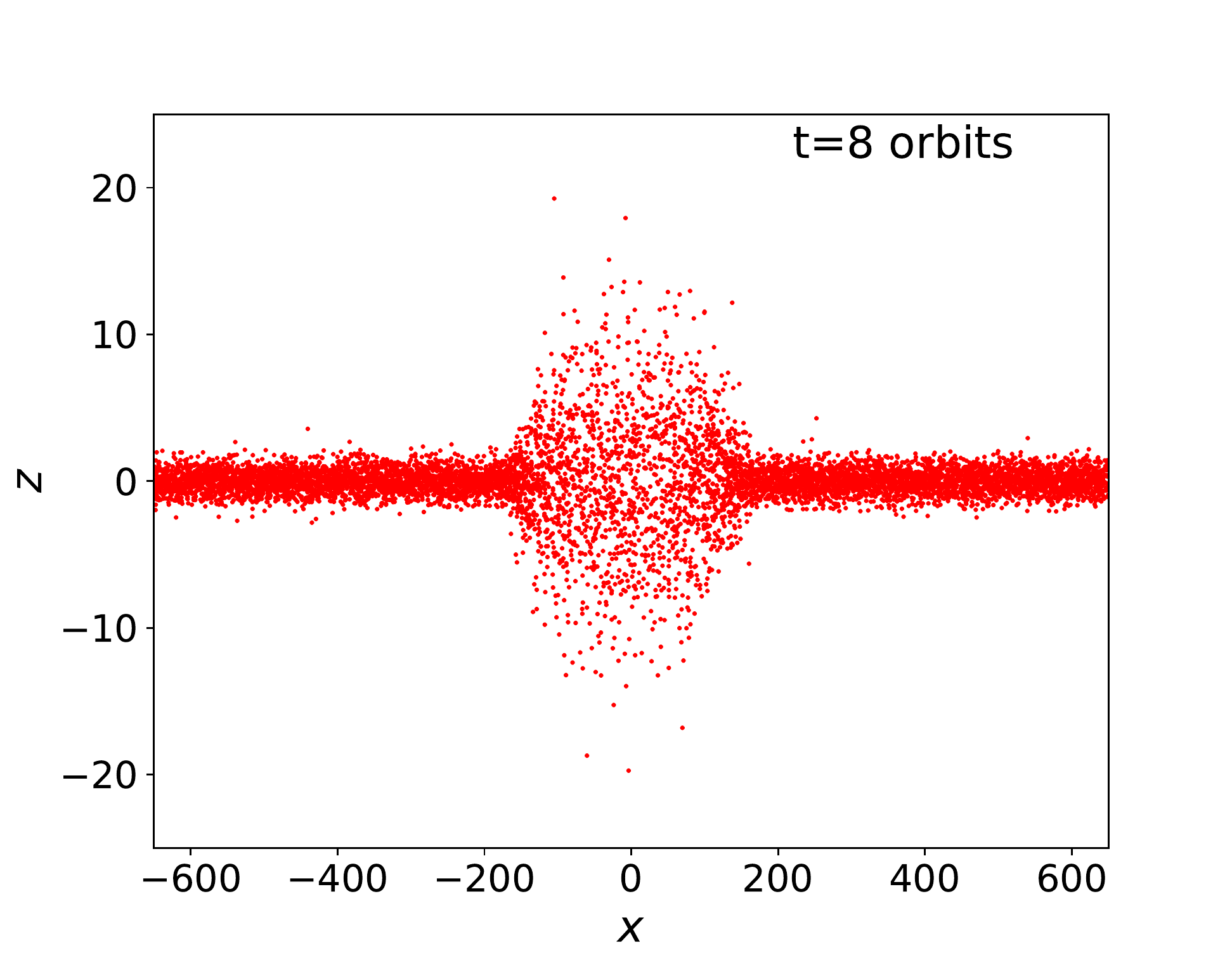}
      \includegraphics[width=0.4\textwidth]{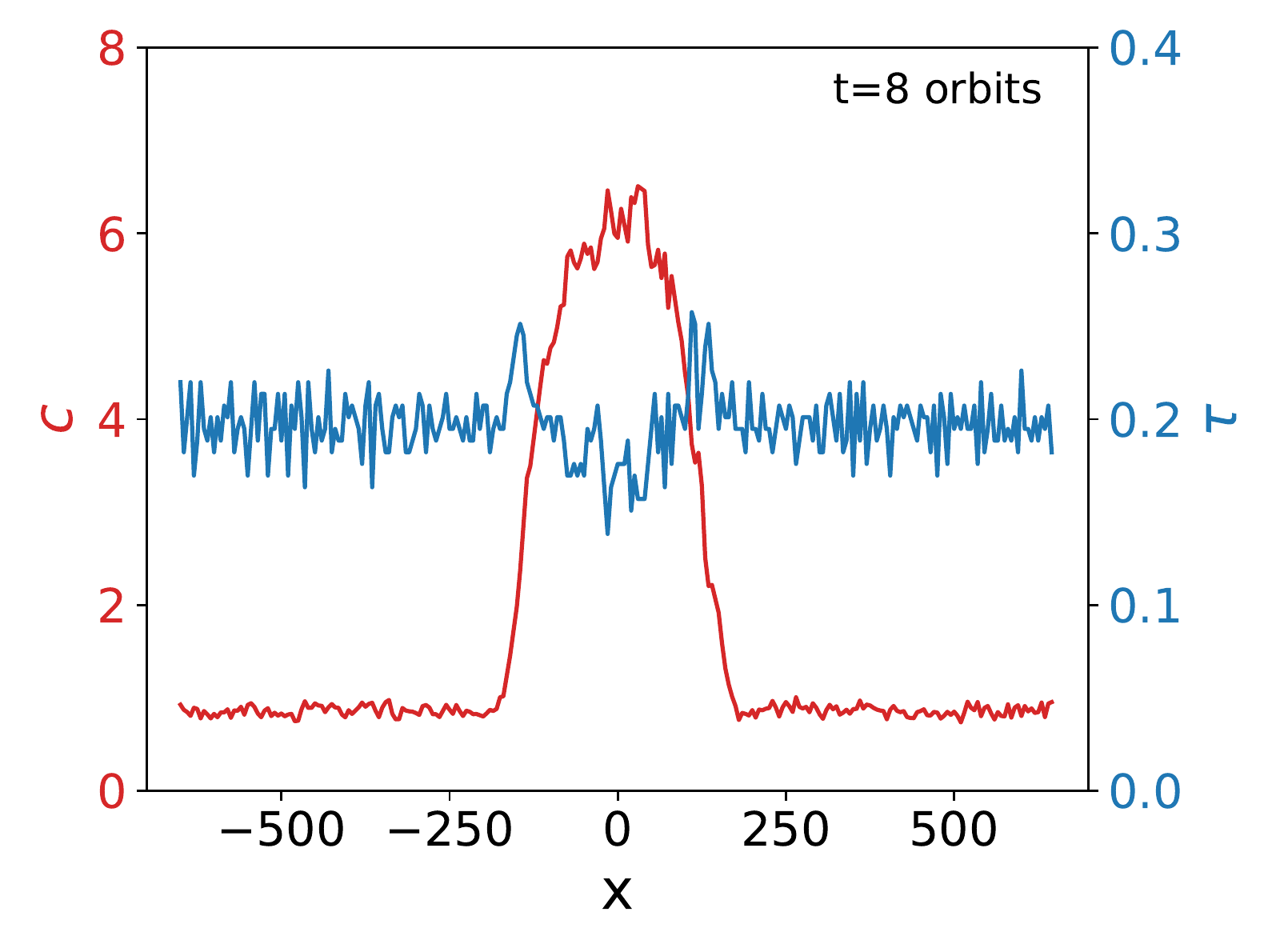}
       \includegraphics[width=0.425\textwidth]{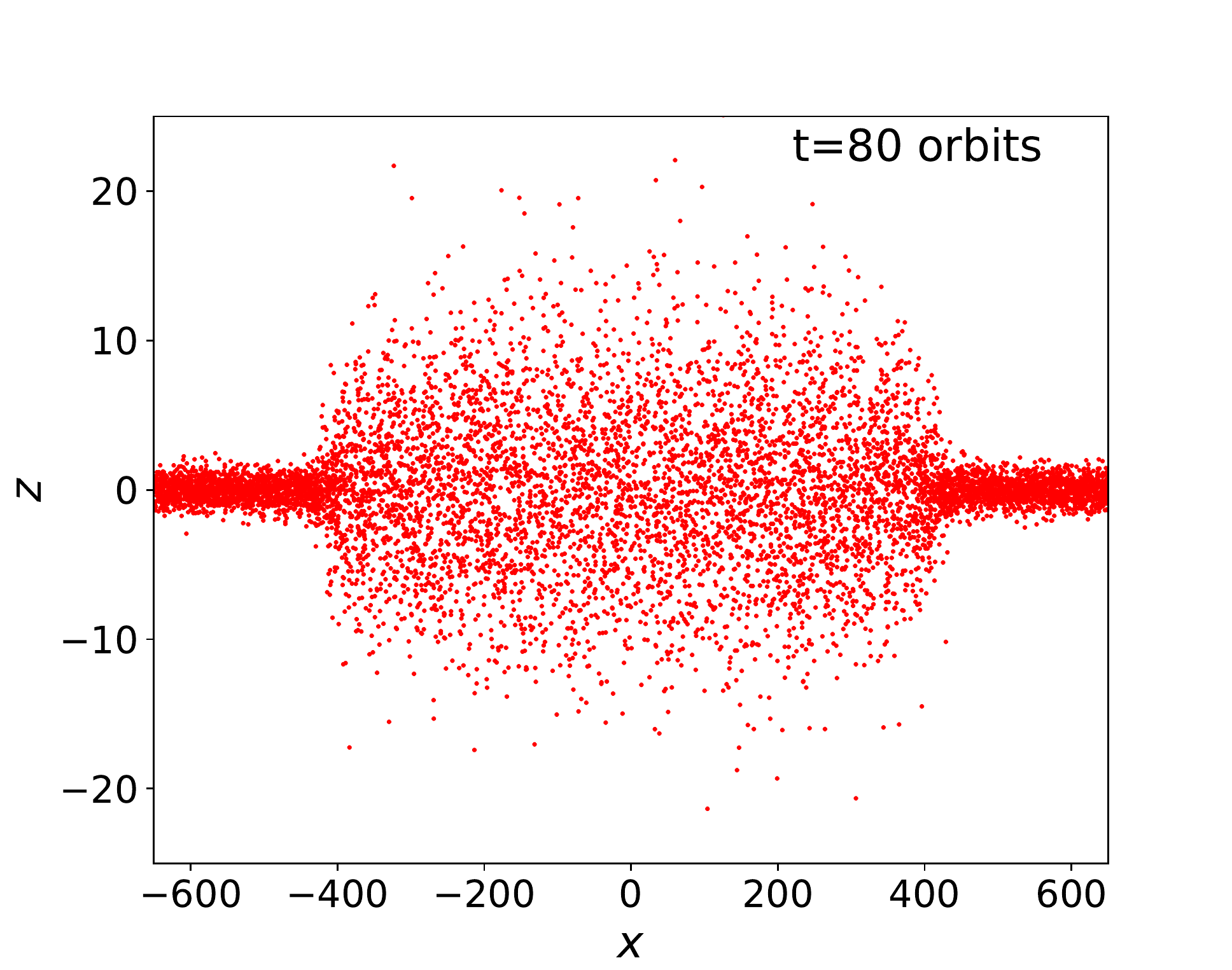}
        \includegraphics[width=0.4\textwidth]{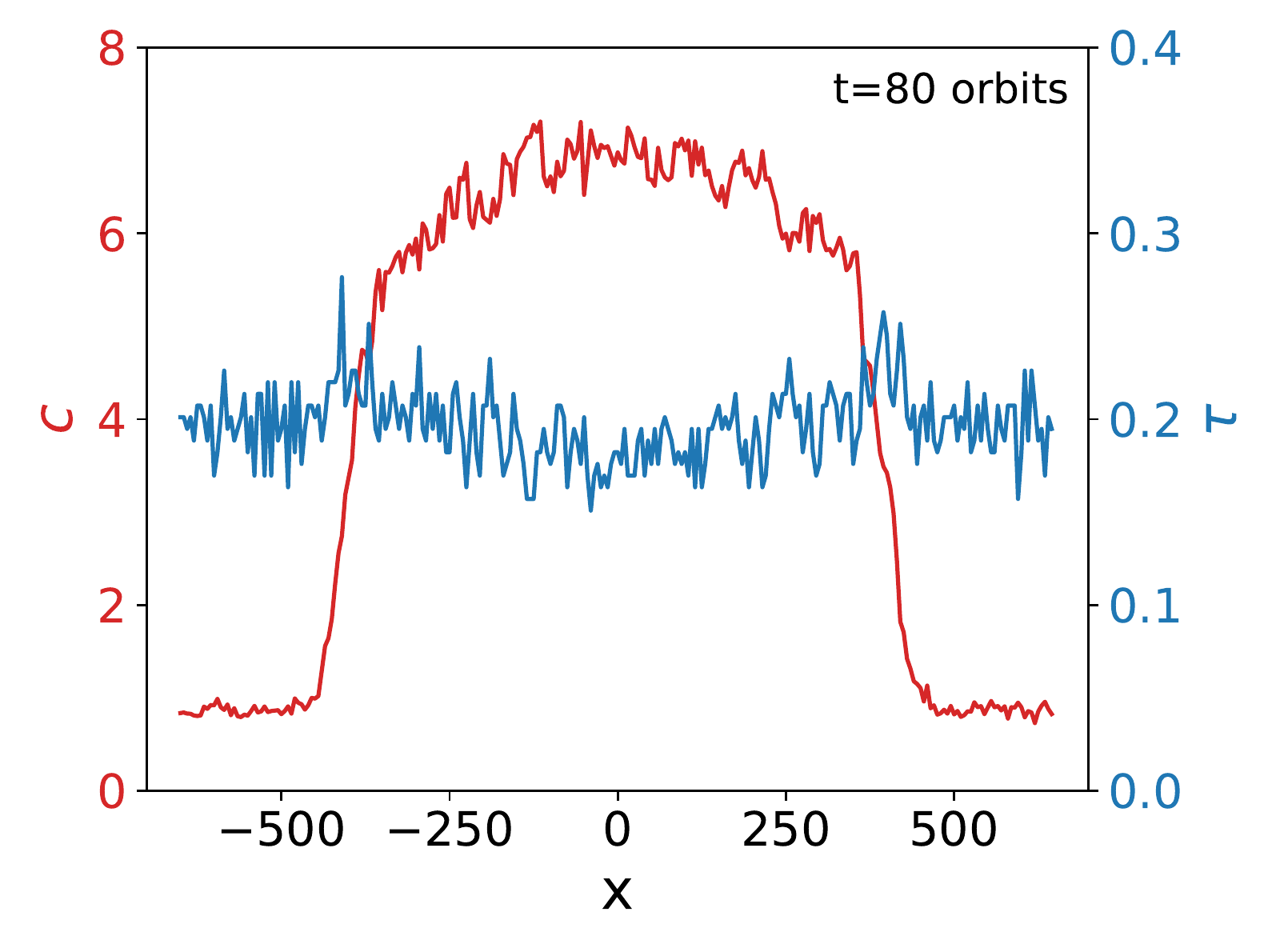}
         \includegraphics[width=0.425\textwidth]{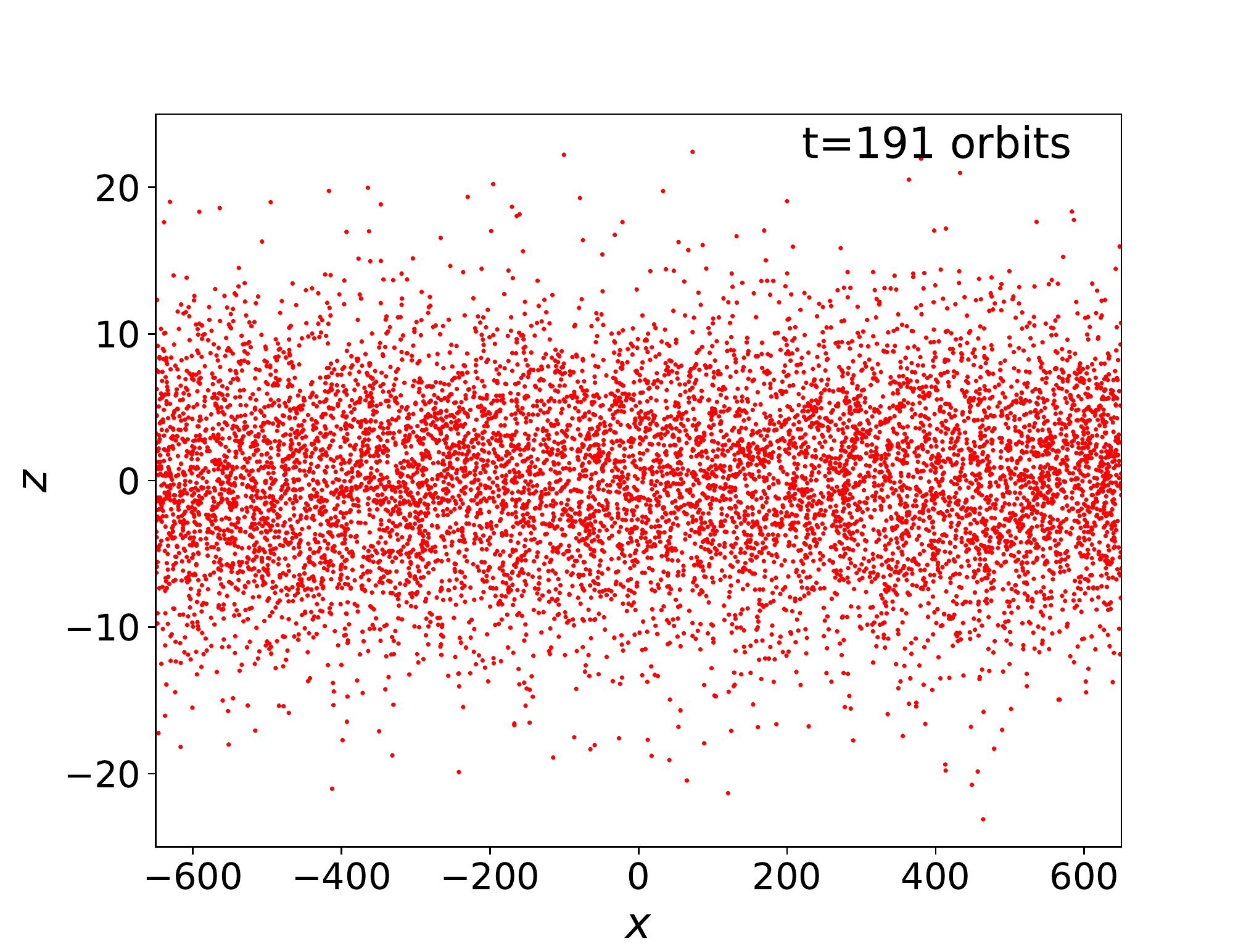}
         \includegraphics[width=0.4\textwidth]{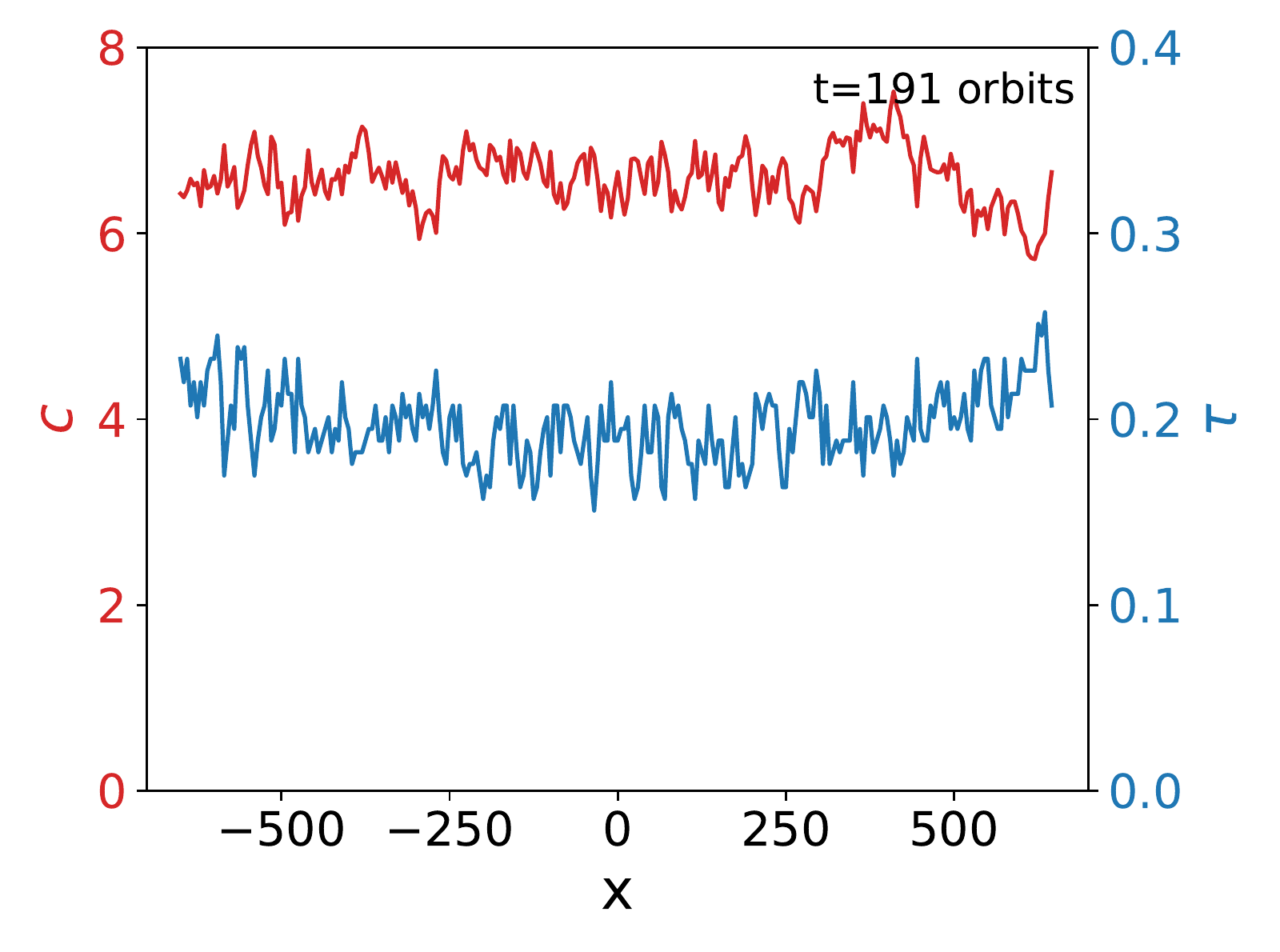}
         \caption{Snapshots of a thermal front at $t=0.8, 8, 80 $ and 191 orbits. Panels on the left describe a projection of ring particles on to the $(x,z)$ plane. Panels on the right depict the $x$-dependent velocity dispersion $c$ (red) and optical depth $\tau$ (blue). }
\end{figure*}

\begin{figure}

\centering
\includegraphics[width=0.5\textwidth]{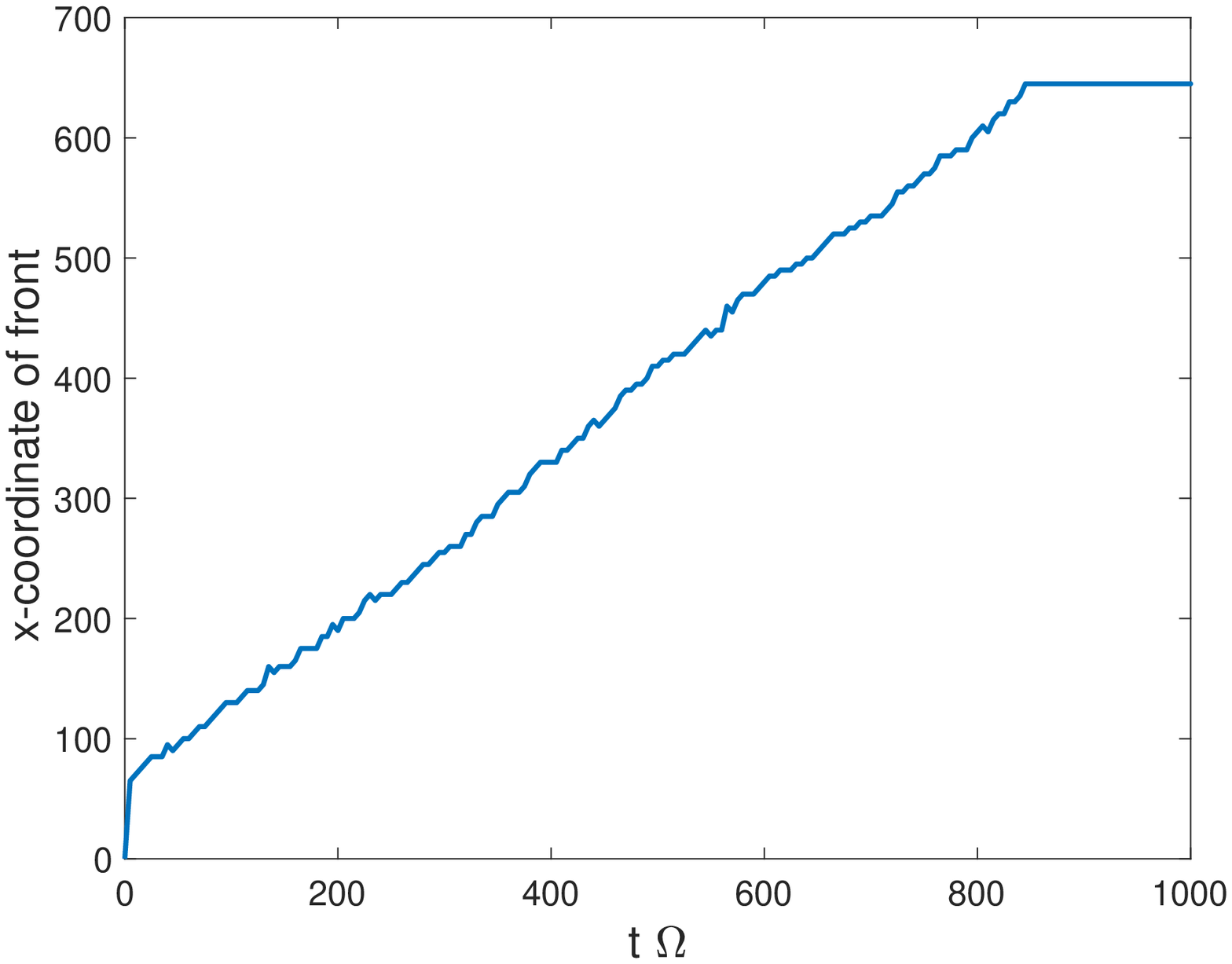}
\caption{Outer front radial location as a function of time in the simulation shown in Fig.~10.}
\label{front speed}
\end{figure}

\subsection{Thermal fronts}

In order to explore the structure and dynamics of fronts connecting equilibria of different temperatures but the same surface density, we concentrate on a single parameter set. The behaviour obtained is then interpreted using a simple continuum model, before other parameters are trialled.

\subsubsection{Fiducial case}

Our fiducial run employs a realistic $\epsilon$ law with the following parameters: $\epsilon_\text{max}=0.75$, $v_\text{crit} = 5$, and $b=1$. We examine a hot and cold state of the same $\tau = 0.2$, with the former possessing $c = 6.7$ and the latter $c = 0.87$. We adopt a wide box of radial size $1000a$ and insert a strip of particles from the (previously computed) hot state in the centre (with radial extent $100a$), while distributing particles from the cold state throughout the rest of the numerical domain. Figure 9 plots this initial condition as a projection of the particle locations in the $(x,z)$ plane. Away from the borders of the hot/cold zones, the ring is in thermal equilibrium.

The subsequent evolution of the ring is shown in Fig.~10, which presents four snapshots at different times on each row. The left panels describe the $(x,z)$ projections of the particles, while the right panels plot the radial variation of $\tau$ (blue) and $c$ (red). As is clear, the two fronts move radially into the cold state, until the hot state takes over the box entirely. Meanwhile,  $\tau$ remain roughly constant throughout, except for some minor deviations around the front itself.

The front speed is constant until the moment that the cold state evaporates. This is demonstrated in Figure 11, which plots the location of the rightmost front as a function of time. A $c$ intermediate between the $c$ in the hot and cold states was selected (here $c=4$) and its $x$ location was determined at each time-step, which provided a means to capture the movement of the front as a whole.  
The front speed is $0.685a\Omega$, thus slightly less than $c$ in the cold state. 

Generally, in bistable systems, the conductivity controls the structure of fronts; a small conductivity yields a narrow transition, while a large conductivity gives a more diffuse transition (e.g.\ Latter and Balbus 2012). In our granular gas, the thermal conductivity $\kappa$ depends on $c$, and thus jumps by at least an order of magnitude as we go from the cold to the hot state (see Table I). This explains why the front structure is sharp near the cold state (though always longer than the `granularity scale', $a$), while broader and smoother near the hot state. The overall width of the front ($\gtrsim 100a$) is hence determined approximately by $\kappa$ in the hot phase.

\subsubsection{Physics of front motion; a simple continuum model}

The basic mechanism driving the movement of a thermal front relies on the finite-amplitude perturbations arising from the proximity of the different states. These perturbations can only be communicated via thermal diffusion.
For example, near a front, the cold state will receive thermal energy (via diffusion) from the adjacent hot state. If the energy received is sufficient to push the cold ring material out of the cold state's basin of attraction, then one might expect it to heat up and settle on the hot state; as a consequence, the front advances into the cold phase. But, by the same token, on the other side of the front, material in the hot state will also be perturbed by the heat flux and will cool down. If this cooled material is pushed beyond the hot state's basin of attraction, then it will undergo a runaway cooling and then we might expect the front to advance into the hot state. Which thermal runaway is favoured on average depends on the relative sizes of the hot and cold state's basins of attraction, which can be approximated (roughly) by how close the intermediate unstable state is to either state (see discussion in the section on metastability, and also Latter and Balbus 2012). 

These ideas can be illustrated by a continuum model. The energy equation of the gas may be written as
$$ \d_t E = \Lambda(E) +\d_x(k \d_x E),$$
where $E=(3/2)c^2$, $\Lambda$ combines viscous heating and collisional cooling, and $k$ is thermal diffusivity ($=2\kappa/(3\sigma)$). Thus $\Lambda=0$ when $E$ is equal to the stable hot, cold, and unstable intermediate steady states, $E_H$, $E_C$, and $E_I$, respectively. Moreover, $d\Lambda/dE <0$ when $E=E_H$ or $E_C$. We assume a steady front, moving at speed $v_f$, with the hot state to the right and the cold state to the left, and thus introduce the comoving variable $\xi=x-v_f t$, which transforms the energy equation into a type of Stefan problem for the front shape $E(\xi)$ and speed $v_f$,
\begin{equation} \label{stefan}
 \d_\xi(k\d_\xi E) +v_f \d_\xi E +\Lambda(E)=0.
 \end{equation}
The boundary conditions are $E\to E_\text{H}$ as $\xi\to\infty$ and $E\to E_\text{C}$ as $\xi\to -\infty$ (hot to the right and cold to the left). This is a nonlinear eigenvalue problem that, after specifying the functional forms of $\Lambda(E)$ and $k(E)$, would normally require a numerical solution. In Appendix B we adopt simple prescriptions for these functions and solve the equation, thereby illustrating some of the main features discussed below and qualitatively reproducing our $N$-body results.

An illuminating expression for the speed $v_f$ can be obtained by multiplying Eq.~\eqref{stefan} by $dE/d\xi$ and integrating between $-\infty$ and $\infty$. After some manipulation, one gets
\begin{equation} \label{vf}
 v_f = -\frac{
\int_{E_C}^{E_H} \Lambda\, dE}{\int_{-\infty}^\infty(dE/d\xi)^2 d\xi}
-\frac{\int_{-\infty}^{\infty}(dk/dE)(dE/d\xi)^3 d\xi}{2\int_{-\infty}^\infty(dE/d\xi)^2 d\xi }.
\end{equation}
If the thermal diffusivity is a constant, the second term is zero. In this case, the sign of $v_f$ is determined solely by the integral of the heating/cooling term $\Lambda$. Because $\Lambda(E_H)=\Lambda(E_I)=\Lambda(E_C)=0$, the integral can be subdivided into (a) a positive part (between $E_C$ and $E_I$) that measures the `size' of the cold state's basin of attraction, and (b) a negative part (between $E_I$ and $E_H$) that measures the hot state's basin of attraction. The proximity of $E_I$ to either $E_C$ or $E_H$ indicates the basins' relative sizes. If $E_I$ is closer to $E_C$, then the integral is dominated by the positive area, $v_f<0$, and the front moves into the cold state. Physically, cold ring material near a front finds it easier to undergo a heating runaway, when perturbed by the front, than hot material finds a cooling runaway; thus, the front advances into the cold material. If $E_I$ is closer to $E_H$, then the converse holds and the front moves into the hot state. Turning now to the top row of Fig. 3 (first panel especially), one naively expects that at low $\tau$ fronts initially move into the cold state, but at higher $\tau$ fronts are slower and then at some critical $\tau$ may reverse direction.

If $k$ depends on $E$ then things are more complicated. The second term in Eq.~\eqref{vf} is a weighted average of $d k/dE$, and shows that a non-uniformity in the transport of heat moderates the effect discussed above. If the front shape is monotonic in $\xi$, then $dE/d\xi>0$ throughout and the sign of the second term is determined by $dk/dE$. As demonstrated in Section 3.2.3 and Table I, $dk/dE>0$, and so the second term in Eq.~\eqref{vf} is always positive, thus biasing the front's movement into the cold state. The underlying mechanism here rests not on the system's bistability but on exacerbating the imbalance in the heat flux throughout the front structure: at any given point more heat is arriving from the hot state than is being evacuated.     

The discussion above suggests that the sharp region at the foot of the front controls the front speed. Taking an order of magnitude approach and equating the three terms in Eq.~(\ref{stefan}) yields the estimate $v_f\sim\sqrt{k_C/t_{th}}$, where the thermal timescale is defined as $t_{th}= E/\Lambda\sim c^2/(\nu\Omega^2)$, and $k_C$ is the diffusivity evaluated in the cold state. Putting in values for the cold state gives us $v_f\sim a\Omega$, which is consistent with the value calculated numerically. The width $\lambda$ of the front extending through the hot phase can then be estimated by balancing the first two terms in Eq.~(\ref{stefan}); we find $\lambda\sim k_H/v_f \gtrsim 500 a$, which is also consistent with the simulation.

\subsubsection{Front stability}

We conducted a short survey of fronts at different $\tau$ and calculated their speeds. When $\tau=0.1$ we found $v_f=0.518$, and when $\tau=0.3$, $v_f= 0.591$. While no clear trend could be observed between $\tau=0.1-0.3$, we expected at larger $\tau$, as we approached the saddle node, that the front speed should decrease. In fact, what we found for $\tau=0.4$ or larger is that the front would slow to a halt and then viscously reshape; i.e. $\tau$ would evolve away from a uniform profile. Ultimately, the system moves to a state of constant angular momentum flux $\tau\nu$, and the thermal front dissolves.

As mentioned earlier, the issue here is that across a thermal front $\tau$ is constant, but $\tau\nu$ is not. As a consequence, mass can potentially build-up/evacuate. If the front moves faster than $\tau$ can be viscously redistributed, then we expect the front to remain coherent and to travel unimpeded. If the front speed is too slow, then it will be viscously reshaped and will collapse. For the model chosen, $\tau\leq 0.3$ corresponds to the first case, and $\tau>0.3$ to the latter. 

A rough criterion for the `stability' of the front to viscous redistribution would tension the relative sizes of the front speed $v_f$ and the viscous diffusion speed. To determine an estimate on the latter, we employ the lengthscale of the abrupt transition at the foot of the structure and thus estimate the diffusion speed as $\sim (\nu_C/\kappa_C)v_f$. A simple criterion for front dissolution requires that this speed is greater than $v_f$, and hence depends solely on the size of the Prandtl number $\text{Pr}=\nu/\kappa$ in the cold state: when Pr is greater than a critical value $\text{Pr}_c$, we expect the front to dissolve. Indeed, Pr increases monotonically between $\tau=0.1$ and $0.4$, though takes relatively small values. At $\tau=0.4$, we find that $\text{Pr} \sim 0.04$, which must be near Pr$_c$.

\begin{figure*}
    \centering
    \includegraphics[width=0.4\textwidth]{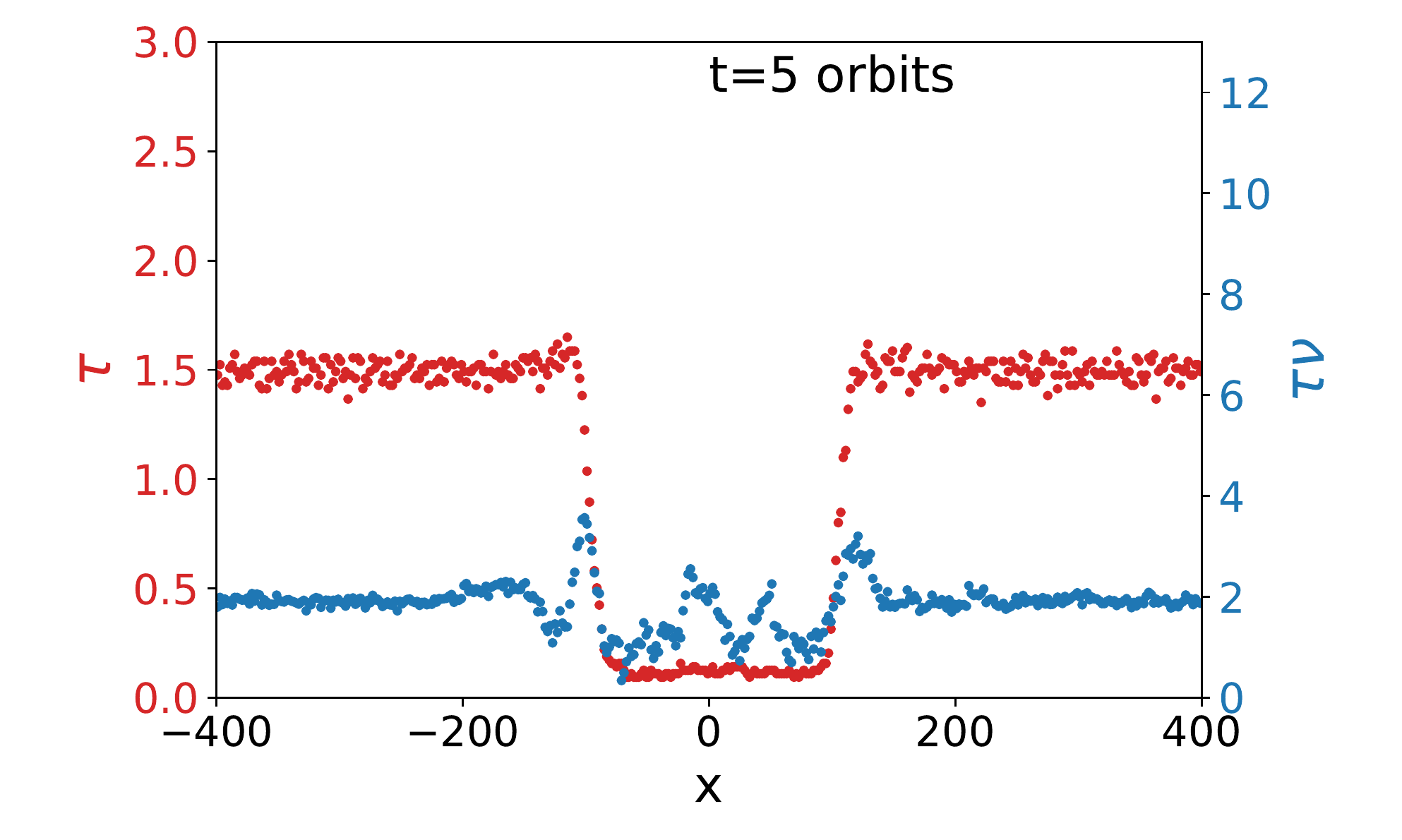}
     \includegraphics[width=0.4\textwidth]{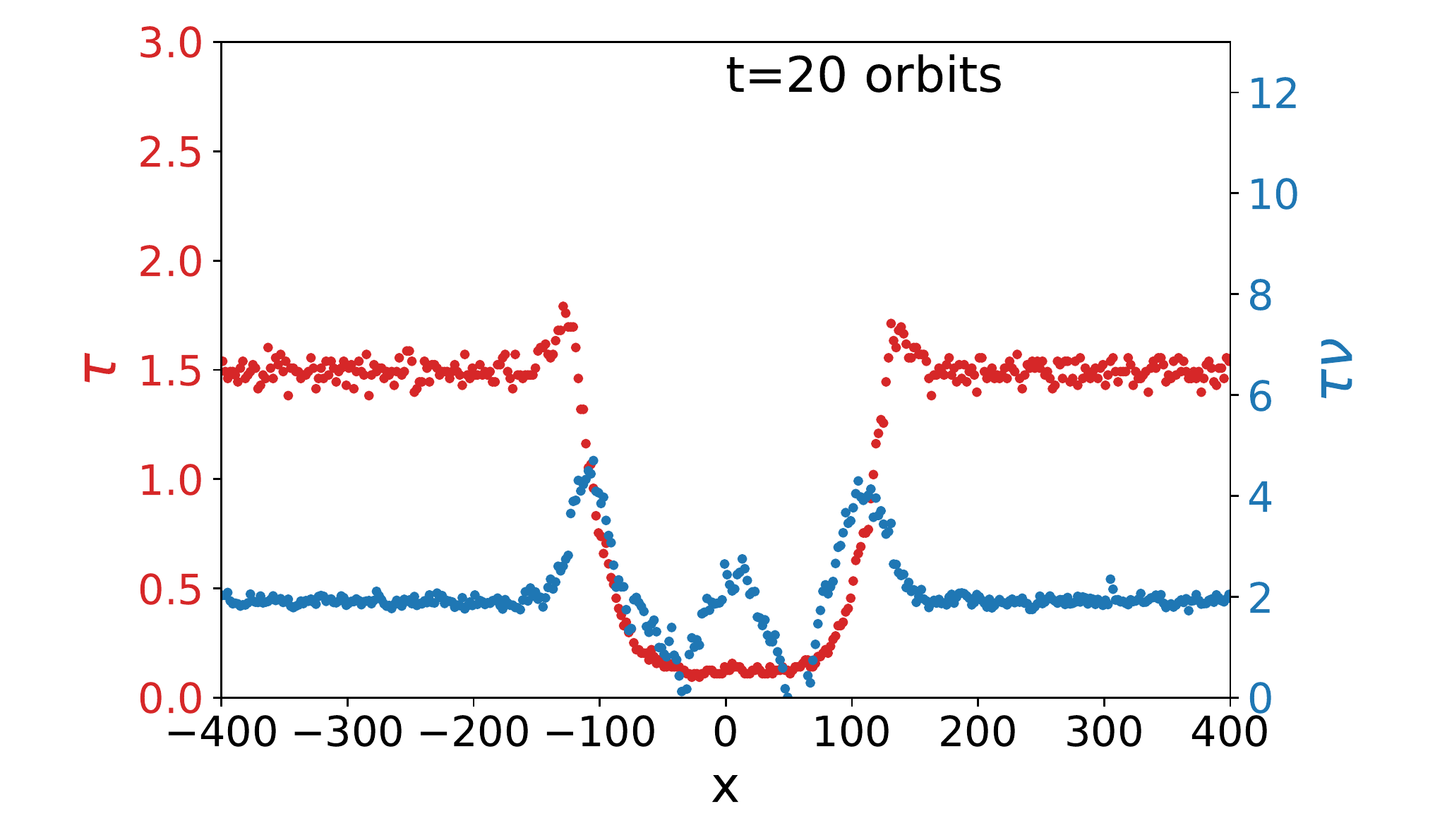}
    \includegraphics[width=0.4\textwidth]{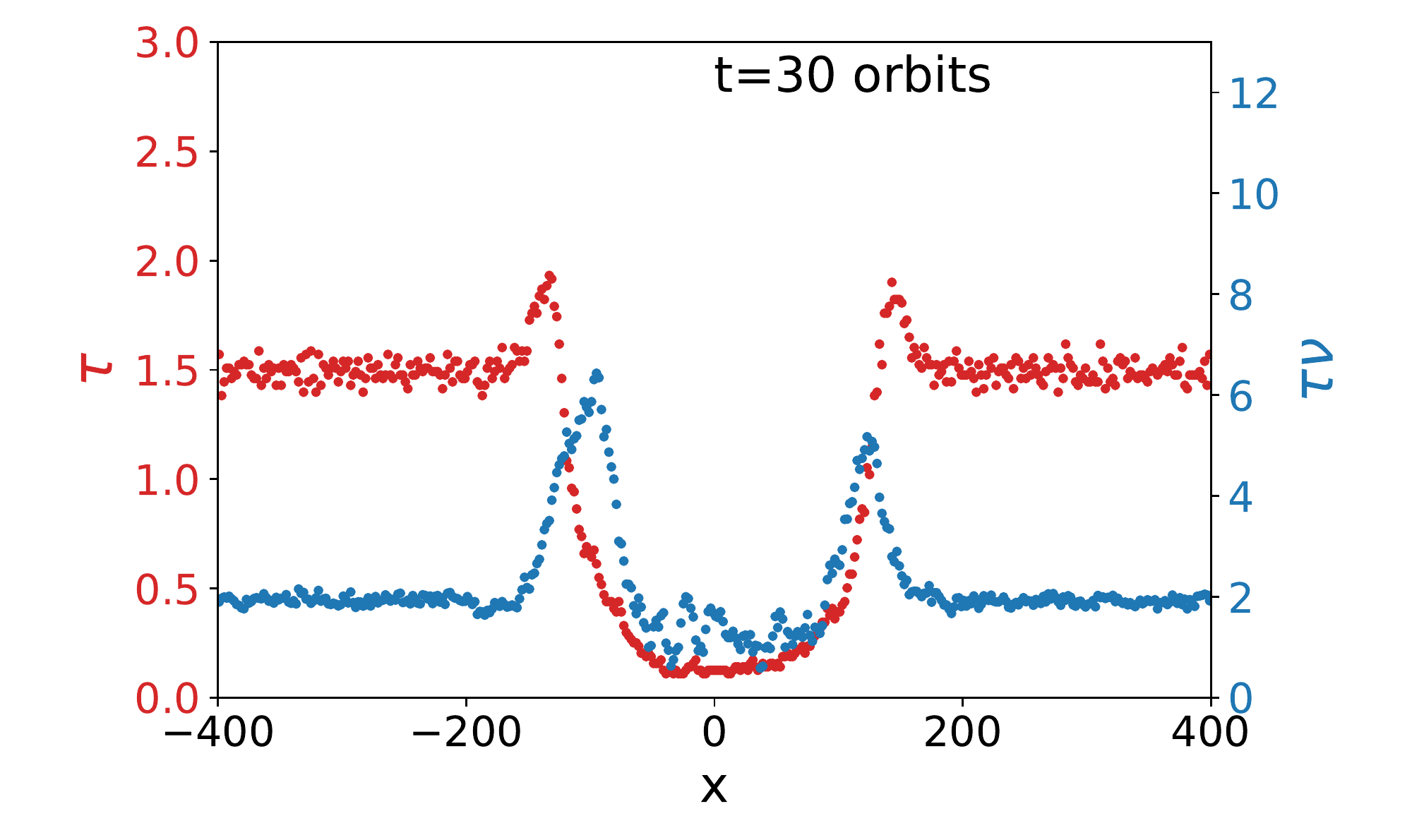}
    \includegraphics[width=0.4\textwidth]{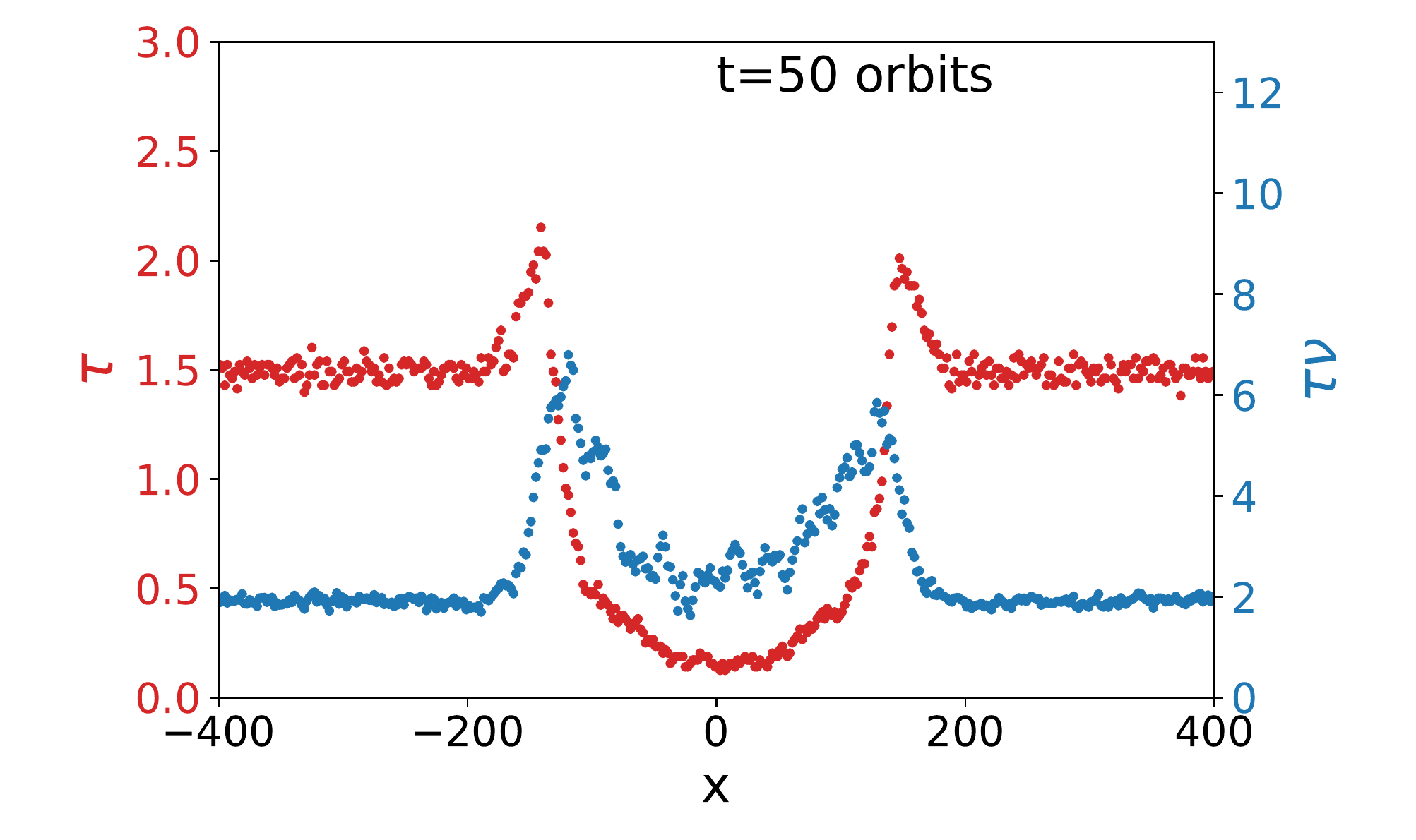}
    \includegraphics[width=0.4\textwidth]{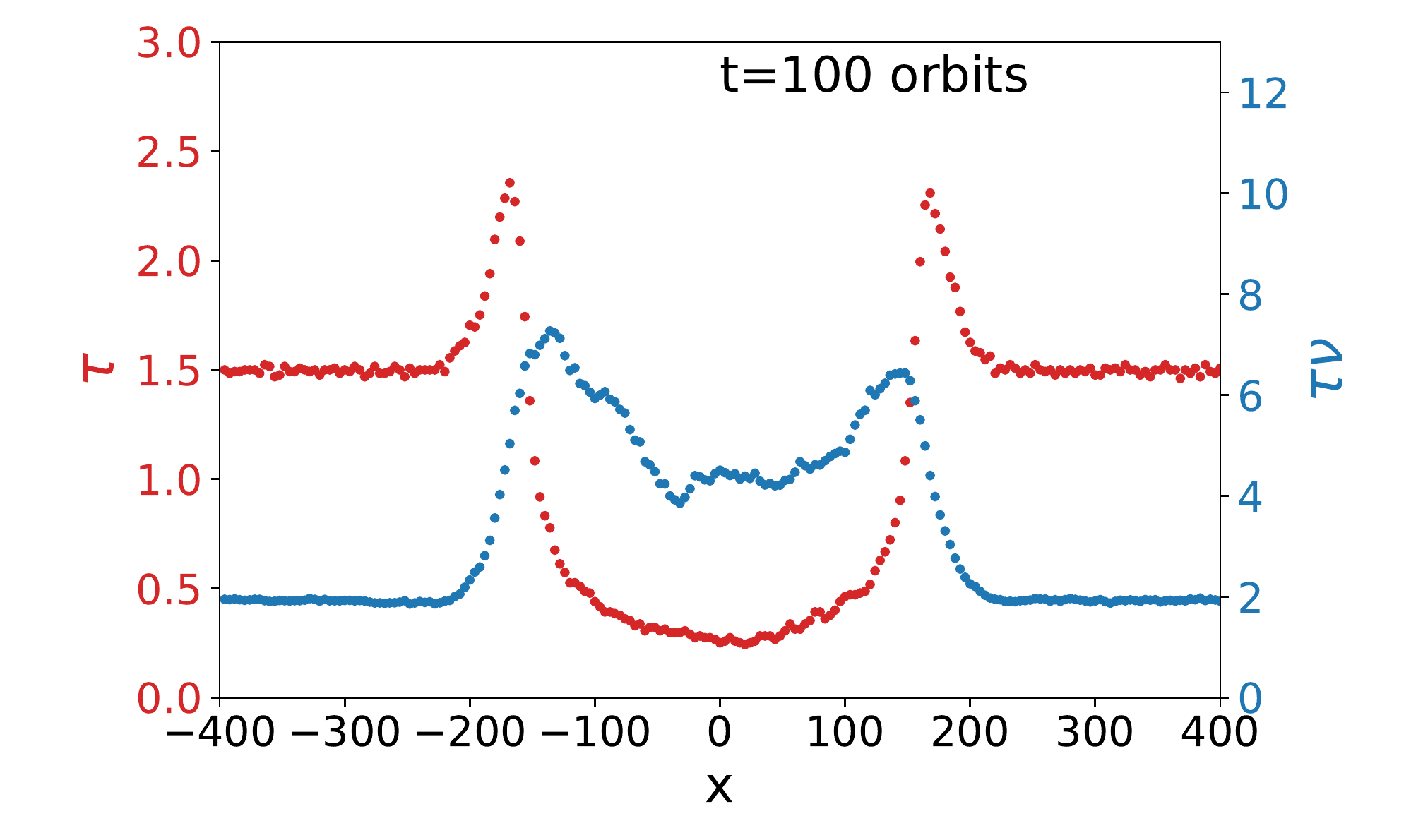}
    \includegraphics[width=0.4\textwidth]{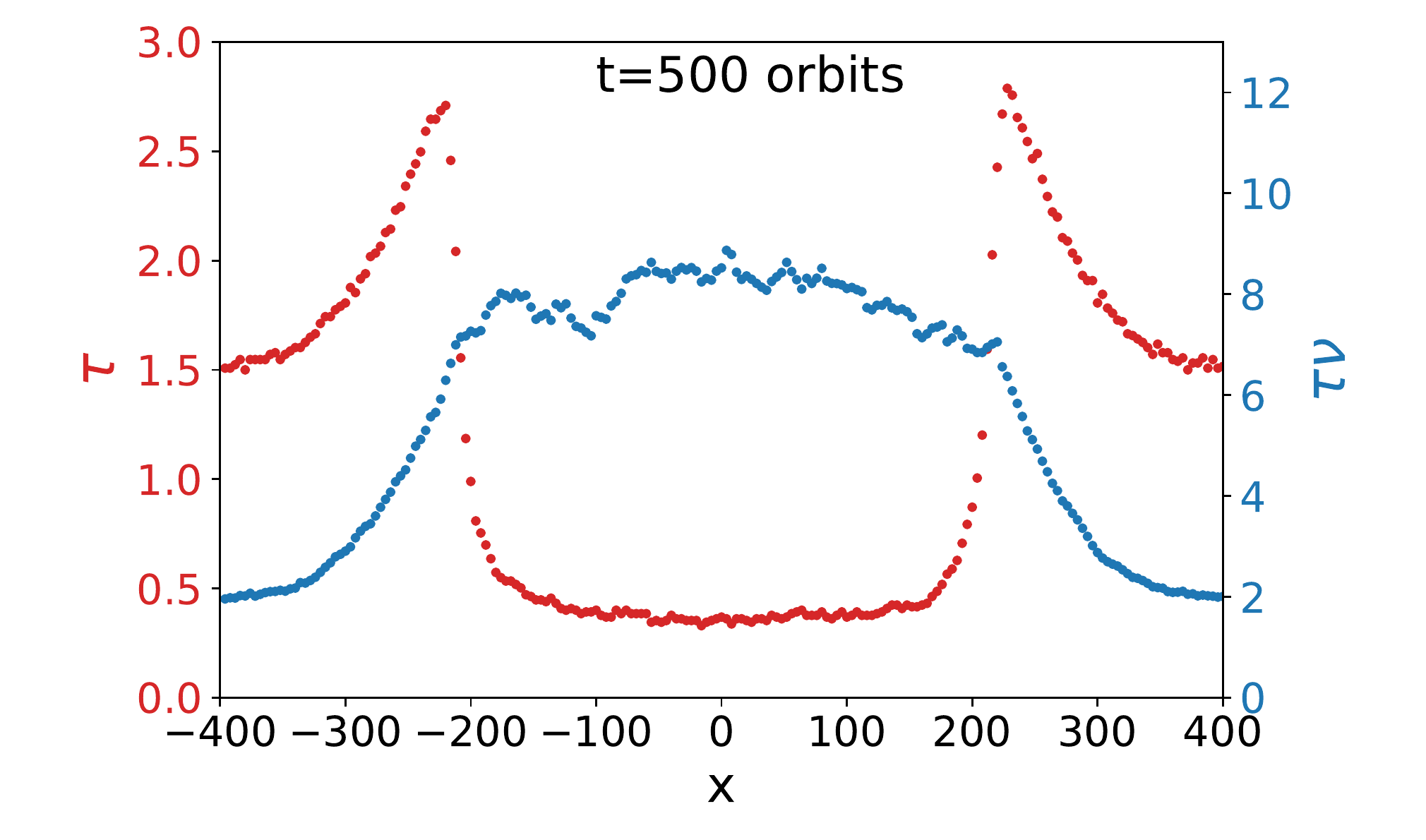}
    \includegraphics[width=0.4\textwidth]{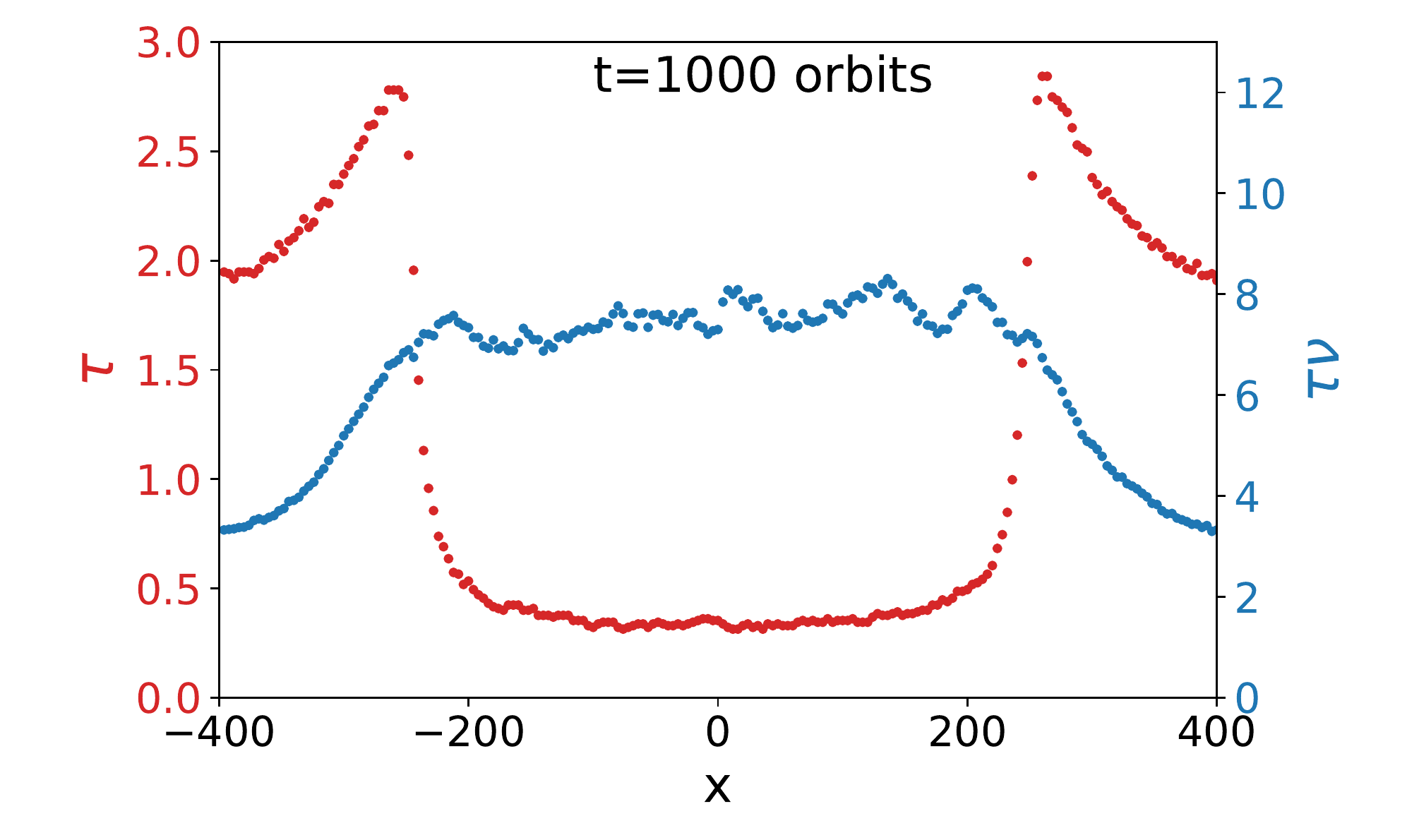}
    \includegraphics[width=0.4\textwidth]{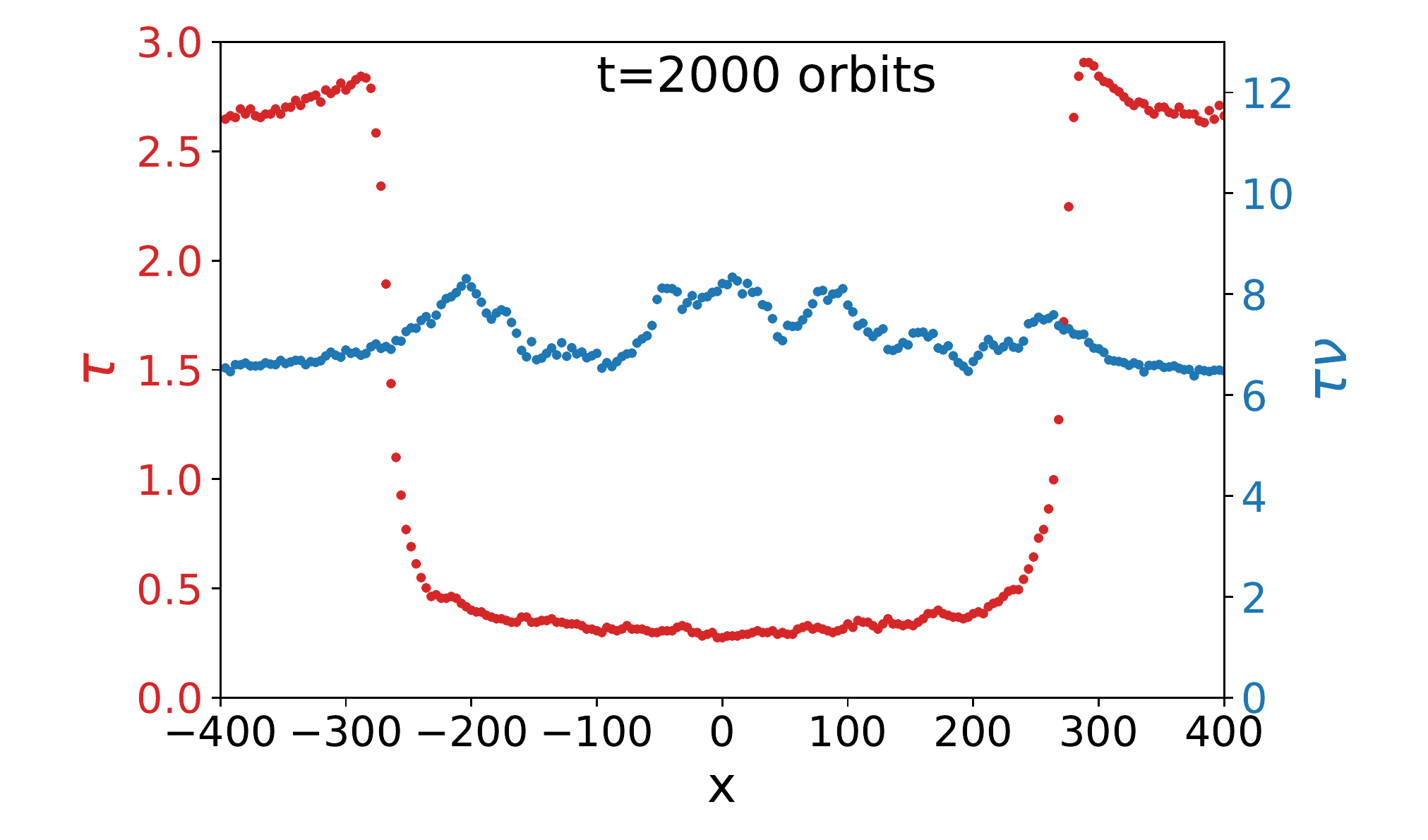}
             \caption{ Snapshots of an example viscous front, showing optical depth and angular momentum flux as a function of $x$. The initial condition connects two states of different $\tau$ but the same angular momentum flux $\tau\nu$. Despite this balance, the system evolves, redistributing mass and angular momentum until a steady state is achieved characterised by a different constant $\tau\nu$. The collision law employs the realistic model with $v_\text{crit}=5$, $\epsilon_{max}=0.923$, $b=1$. Snapshots are at $t=5,20,30,50,100,500,1000$, and $2000$ orbits. }
\end{figure*}

\subsection{Viscous fronts and viscous instability}

Given the issue of the unbalanced angular momentum in thermal fronts, it is natural to explore fronts that join states with the same viscous transport properties, specifically $\nu\tau$. We present simulations of such joined states in this subsection, in addition to a short treatment of viscous instability.  

A simple continuum model can guide our expectations. In the shearing sheet, the one-dimensional diffusion equation for viscous Keplerian disks is $$\d_t\tau = 3\d_x^2 (\nu\tau)$$ 
(e.g.\ Lynden-Bell and Pringle 1973). Suppose a viscous front moves with speed $v_f$ with $\tau\to \tau_A$ as $x\to -\infty$ and $\tau\to\tau_B$ as $x\to\infty$. As earlier, we adopt a comoving variable $\xi=x-v_f t$, which permits the complete integration of the problem. We find that $v_f=0$ (the structure must be stationary) and $\nu\tau\,\, (=\nu_A\tau_A=\nu_B\tau_B)$ is a constant throughout the entirety of the front. The last constraint is a potential difficulty: while it is possible to find two homogeneous steady states of the same $\nu\tau$ (cf. panels in the bottom row of Fig.~3), a realistic front will have a finite width in which $\tau$ will vary and thus $\nu\tau$ will deviate from the required constant value. Our simulations show, in fact, that the system can overcome this problem by settling on a front structure in which the \emph{average} $\nu\tau$ equals $\nu_A\tau_A=\nu_B\tau_B$.

\begin{figure*}
    \centering
     \includegraphics[width=0.325\textwidth]{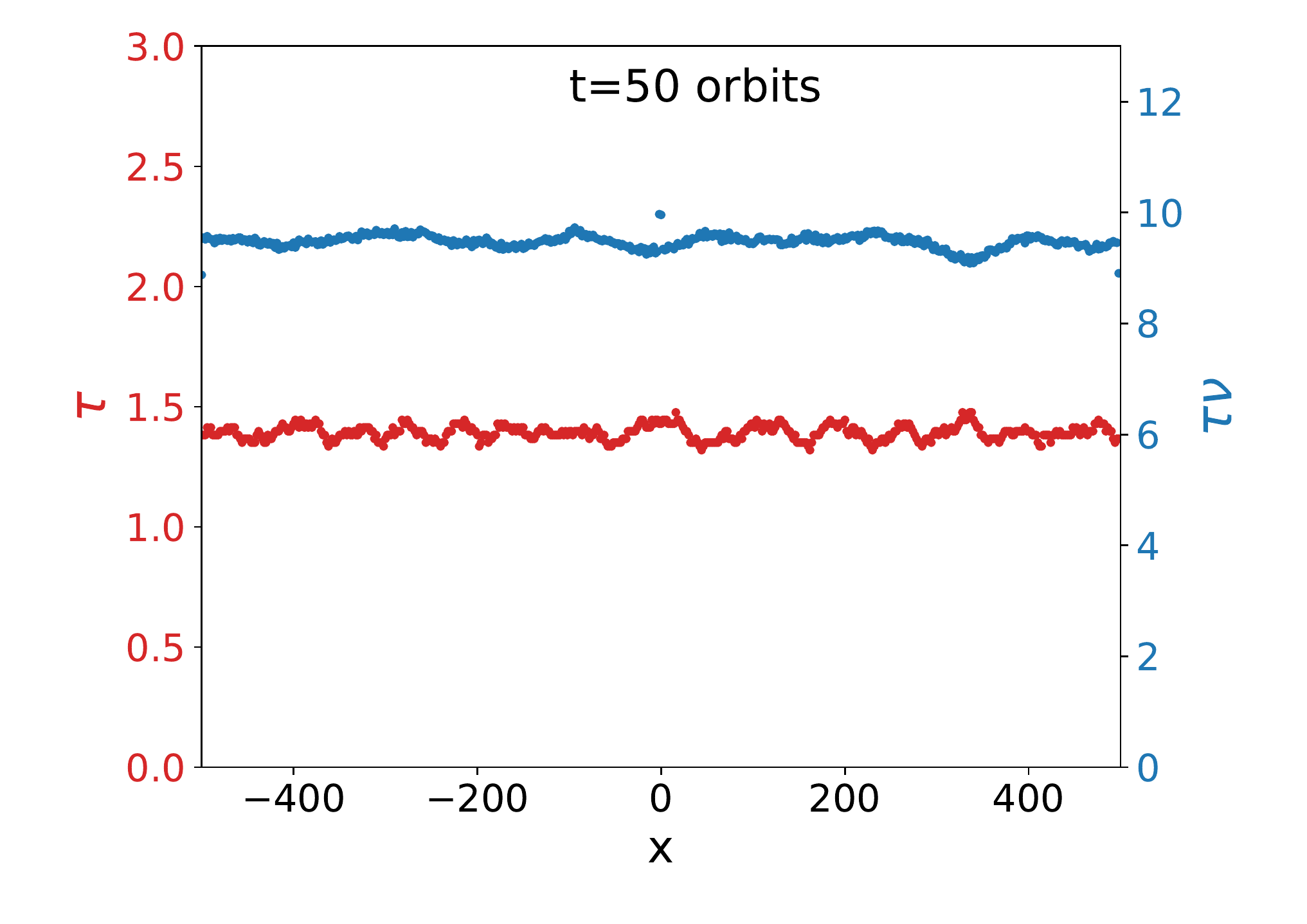}
      \includegraphics[width=0.325\textwidth]{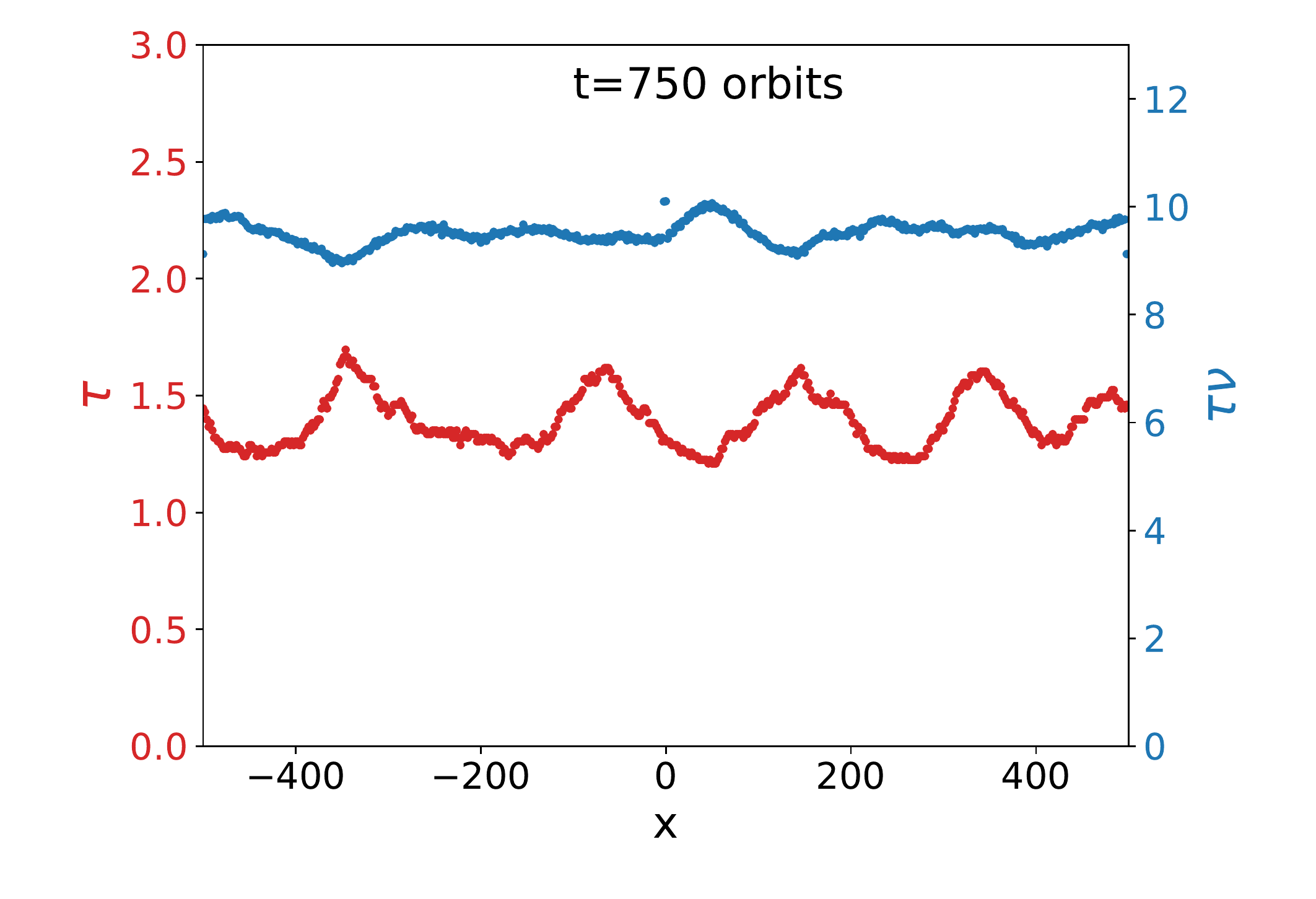}
        \includegraphics[width=0.325\textwidth]{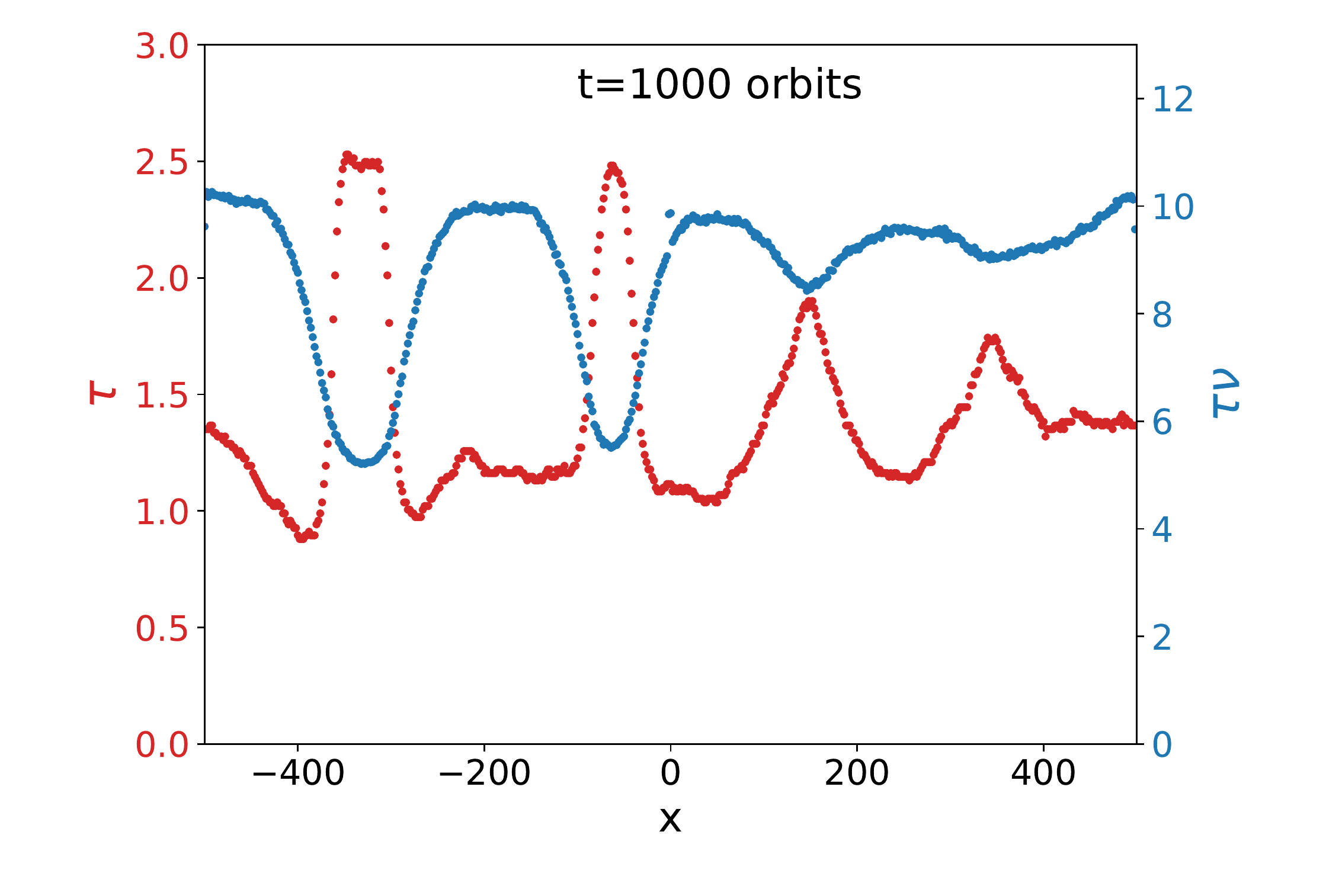}
         \includegraphics[width=0.325\textwidth]{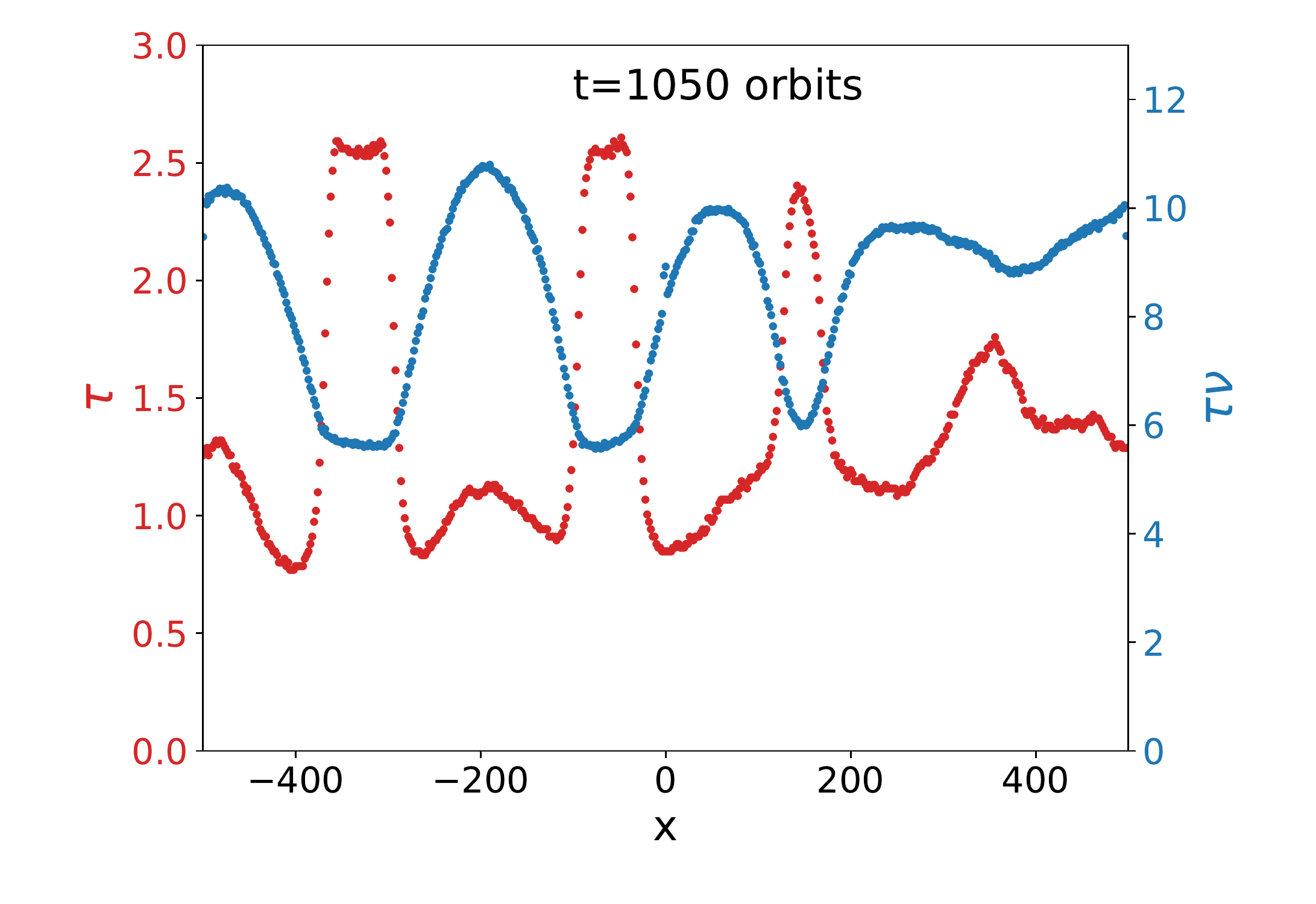}
         \includegraphics[width=0.325\textwidth]{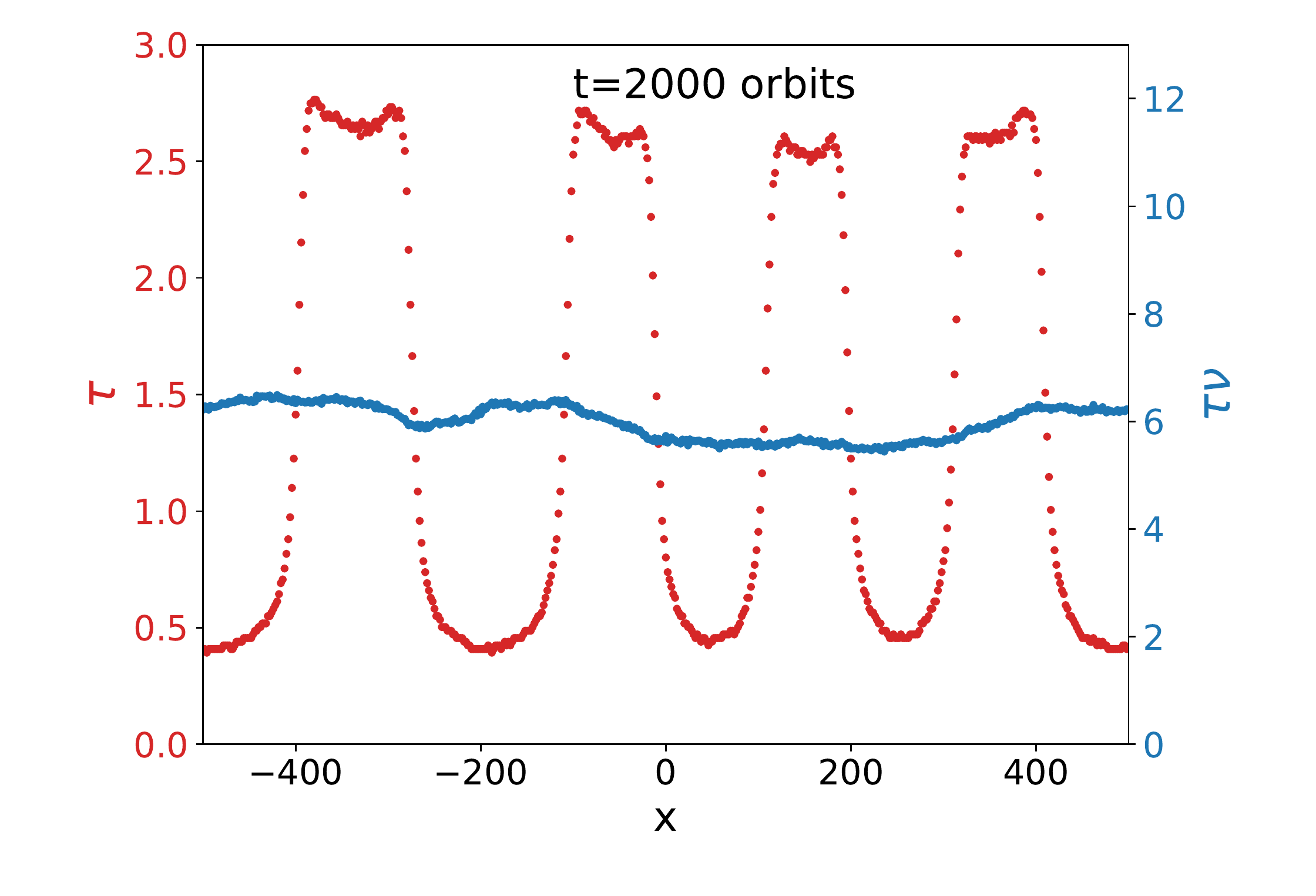}
         \caption{Snapshots showing the progress of viscous instability starting from an unstable state of $\tau=1.4$. The collisional parameters are $v_\text{crit}=5, \epsilon_\text{max}=0.923, b=1$. The panels describe the $x$ dependent optical depth $\tau$ (red) and the angular momentum flux $\tau\nu$ (blue). Snapshots are at $50$, $750, 1000, 1050$, and $2000$ orbits.  }
\end{figure*}

\subsubsection{Fronts}

We present a fiducial simulation with the realistic law, and parameters $b=1$, $\epsilon_\text{max}=0.923$, and $v_\text{crit}=5$. To construct a suitable initial condition that might produce a viscous front, we select two thermally and viscously stable states with the same $\nu\tau$ from the bottom right panel of Fig.~3. Such pairs are joined by horizontal lines. We select two states of the same angular momentum flux $\nu\tau\approx 2$, with optical depths $\tau=1.5$ and $\tau=0.16$. The numerical domain is chosen to be sufficiently large ($L=800$) to accommodate relatively undisturbed expanses of the two states, in addition to the front itself; the low $\tau$ state is placed between $x=-100$ and $100$, with the high $\tau$ state taking up the remainder of the box.

Figure 12 shows eight snapshots of the resulting simulation at different times. In each panel we plot $\tau$ (red) and $\tau\nu$ (blue). At $t=0$, the angular momentum flux $\tau\nu$ is a constant, but $\tau$ undergoes two jumps (at $x=\pm 100a$). As the system evolves, the two jumps/fronts relax and exhibit a characteristic width, with $\tau$ taking values between those of the two steady states. An immediate consequence is that the angular momentum flux within the fronts begins to deviate from the fixed value $\approx 2$. In fact, the first four panels show that it takes significantly larger values than 2, in agreement with the bottom right panel of Fig.~3, which shows that states with $\tau$ between $0.16$ and $1.5$ exhibit $\nu\tau > 2$. Because of the enhanced flux in the fronts, mass is being transported out of the fronts, which then appear to move as the system evolves far way from the initial condition.

Ultimately, we find that the system redistributes the mass throughout the numerical domain so that $\tau\nu$ is roughly constant ($\approx 7$), but still allows for strong variations in $\tau$. This outcome is not a constant $\tau$ state, but consists of two static viscous fronts joining two homogeneous states of $\tau\approx 0.4$ and $2.7$, which according to Fig.~3 possess the same angular momentum flux ($\sim 7)$. Evidently, the front that joins the two states also possesses a similar approximate flux, though this is difficult to determine from Fig.~3. A similar final state was found by Salo and Schmidt (2010) when simulating the viscous instability directly (see next subsection).

This static structure is an interesting outcome for the system, but we stress that it is possible only because of the periodicity of the numerical domain. Owing to those boundary conditions, mass in the whole domain can be redistributed until the desired constant $\nu\tau$ state can be found. 
 In a more realistic setting, the system is unlikely to come to steady state and the front will continue to move until it encounters large-scale variations in background disk properties, etc.

\subsubsection{Viscous instability}

In the previous subsection we explored two adjoined viscously stable states, but the lower right panel of Fig.~3 indicates that there is a branch of viscously unstable states of intermediate $\tau$ between roughly 0.8 and 1.6. An obvious question is: to where does the system evolve if started from one of these states?  
We thus present a simulation with the same collisional parameters as earlier, but with a homogeneous $\tau$ of 1.4. According to Fig.~3, this state is viscously unstable. Figure 13 shows 5 snapshots of the system's evolution.

Despite possessing a constant $\tau\nu$, the system moves slowly away from this state and begins to develop growing patches of high and low $\tau$. Unlike the previous subsection, where the evolution is being driven by large-scale flux imbalances, here there is an instability mechanism, in which small-scale fluctuations in the flux self-reinforce (Lin and Bodenheimer 1981, Lukkari 1981, Ward 1981). Ultimately, the system settles on a sequence of distinct high-$\tau$ islands surrounded by relatively dilute regions, but both with roughly the same flux ($\approx 6$, in this case), as is necessary for a steady state. 

These results are very similar to those predicted by H\"ameen-Anttila (1982) and witnessed in Lukkari (1981) and Salo and Schmidt (2010), though they use a monotonic collision law. A key difference is that in the monotonic $\epsilon$ simulations, the final outcome joins states from the same branch, while in our non-monotonic simulations states from different branches adjoin. An interesting consequence of this is that it is still possible for the system to separate into a sequence of high and low $\tau$ states (of the same $\nu\tau)$, even when there is no intermediate viscously unstable state. In particular, this appears achievable for the parameters of the middle column in Fig.~3. More generally, systems with non-monotonic collision laws have more freedom to exhibit viscous phase-separation in radius.

\section{Discussion and Conclusion}

Most previous work describing the local collisional dynamics of Saturn's rings uses relatively simple collision models. Given the poorly constrained nature of the collisions, and the numerical challenges involved, this is understandable, and indeed some success has been achieved in certain applications (e.g. self-gravity wakes, viscous overstability). However, current models still fail to describe much (if not most) of the irregular axisymmetric structure exhibited in Saturn's B and C rings. This invites us to experiment with other more complicated collision laws, in particular those that account (in a basic way) for surface regolith on ring particles, which is deemed to be present and important (e.g.\ Nicholson et al.~2008, Morishima et al.~2012, Deau 2015).   

 We conduct $N$-body simulations with the REBOUND code of a local patch of Saturn's rings in which particles undergo collisions with a prescribed coefficient of restitution $\epsilon$ depending on impact speed. The main novelty of our approach is to employ an $\epsilon$ that is a non-monotonic function of impact speed, as is suggested by theoretical and experimental studies of regolith-coated particles (cf.\ Section 2.1). Below a critical impact speed we set $\epsilon=0$, though neglect particle sticking. This relatively minor change in the physical set-up immediately introduces major thermodynamical changes. For the same optical depth, the rings yield two thermally stable steady states, a hot $c\gtrsim 4a\Omega$ and a cold $c< a\Omega$ state. Which is selected depends on the local thermal and/or dynamical history, and thus different ring radii might fall into one or the other. 
 
 An obvious follow up question is to ask what happens at the boundaries of two adjoining different states? We run additional simulations in larger domains and find that in general the hot state will engulf the cold state, with the transition front moving at a speed $\approx 0.5 a\Omega$. Slower moving fronts break down because of the imbalance in angular momentum flux across the transition. Stationary `viscous fronts' are also simulated which join states of different optical depth and $c$ but the same angular momentum flux. Note that it need not necessarily be the case that hot states always take over: smooth variations in the ring's background properties may change propagation, and large amplitude perturbations (meteoroids, density waves, gravity wakes, etc.) will also complicate the picture.    
 
Our simulation results are exploratory, and should be taken as a demonstration of what happens when one relaxes the strong modelling assumptions of previous work. They are perhaps not yet ready for direct application to structure formation in Saturn's rings, not least because of the parameters in our regolith laws are poorly constrained. Nonetheless, it is irresistible to speculate. We anticipate that a thermal front, connecting a warm and cold state of the same dynamical optical depth, gives rise to photometric variation (which the Cassini cameras may have picked up) but no variation detectable by occultation experiments. This is precisely the situation in the C-ring plateaus (Hedman and Nicholson 2013), and indeed, there is evidence of size segregation across these structures which may tie in to the greater chance of sticking in the colder phase (Marouf et al.~2013, Colwell et al.~2018). It may also be relevant for the 10km striations shown by Cassini's cameras in the A and B-rings (cf.\ Figs 5A and 5B in Porco et al.~2005). On the other hand, the steady viscous fronts our simulations support, which connect states of high and moderate optical depth, bear some resemblance to the disjunct bands in the middle B-ring (Colwell et al.~2009). A great deal more theoretical work and modelling is needed before these associations can be made secure. In particular, applications to ring regions exhibiting self-gravity wakes must remain tentative until we produce better constrained estimates on typical sticking speeds.

Other areas of future work could explore the interplay between the hysteresis and self-gravity wakes, on one hand, and viscous overstability, on the other. For example, we might anticipate wakes appear only in the cold state, changing its viscous properties, and providing energy to jump into the hot state. More generally wake activity will produce enhanced heating and thus a change in the thermodynamic balances calculated in this paper. Viscous overstability generates nonlinear travelling wavetrains which may also favour the cold phase; these waves will reflect off the boundaries between states, hence complicating the nonlinear saturation of the wave turbulence. Simulations including realistic photometry of thermal fronts might help establish if they might correspond to any observable structure (Salo and Karjalainen 2003). Finally, the robustness of bistability must be established when particle sticking is permitted, as in recent simulations by Ballouz et al.~(2017) and Lu et al.~(2018).

\section*{Acknowledgments}
The authors thank the reviewer Heikki Salo and Juergen Schmidt, who generously provided a set of helpful and thorough comments that markedly improved the paper. 

\section*{Data Availability}

The data underlying this article will be shared on reasonable request to the corresponding author.

\appendix

\section{Convergence tests}
\label{sec:convergencetests}
We present some results showing the behaviour of a subset of our equilibrium solutions as the numerical parameters are varied. In particular, we explore their dependence on the size of the time-step $dt$ and the numerical domain, showing that convergence is achieved when the former is sufficiently small and the latter sufficiently large. To simplify the study, we adopt a standard Bridges law for two different $v_\text{crit}$ (yielding hot and warm equilibria) and also a constant $\epsilon=0$ (yielding cold equilibria. We examine very dilute cases $\tau=0.1$ and very dense cases $\tau=2.5$, thereby determining the numerical requirements at the physical `boundaries' of our main set of results, and thus for the main results themselves.

\begin{figure}
    \centering
    \includegraphics[width=0.8\columnwidth]{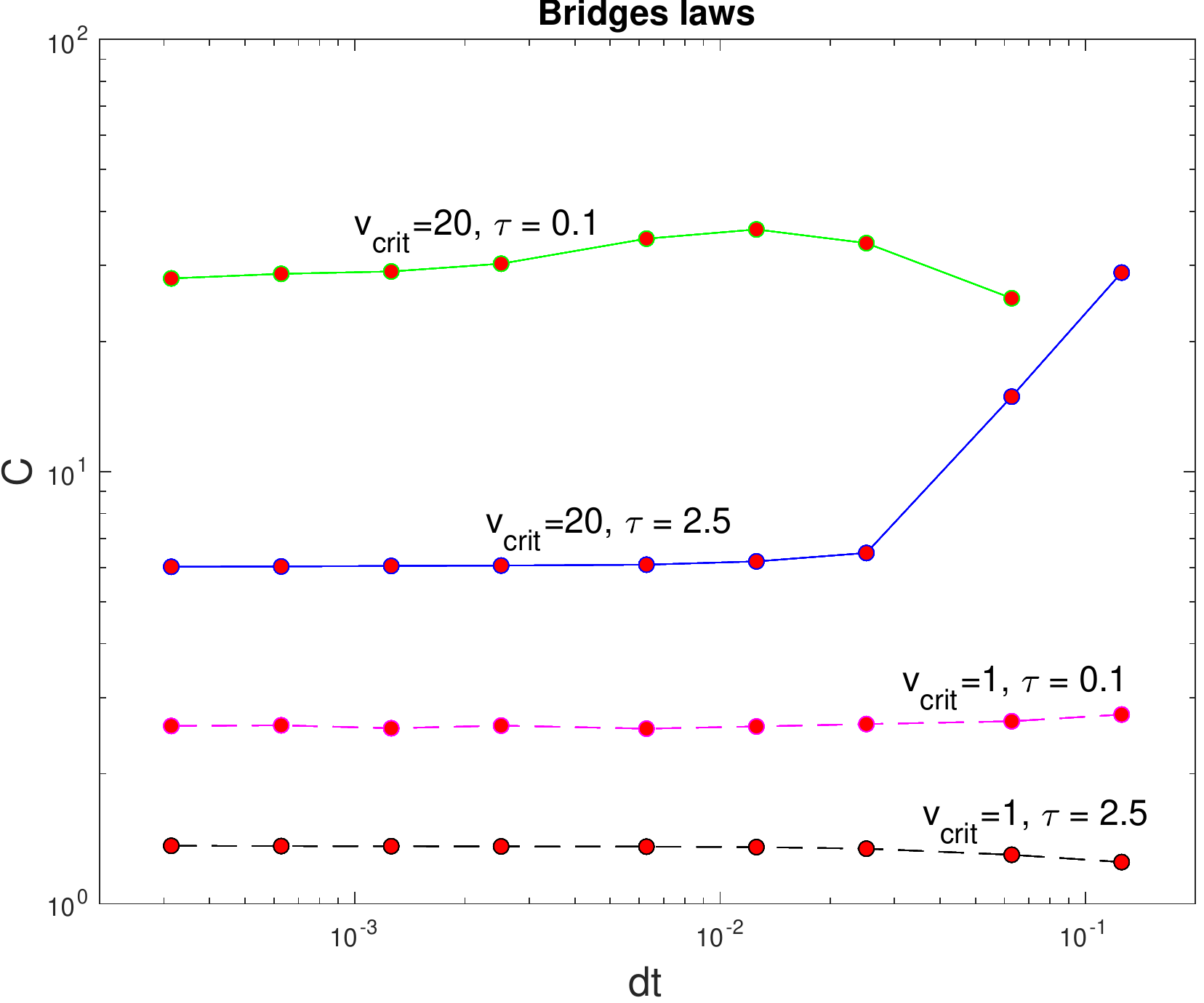}
    \includegraphics[width=0.8\columnwidth]{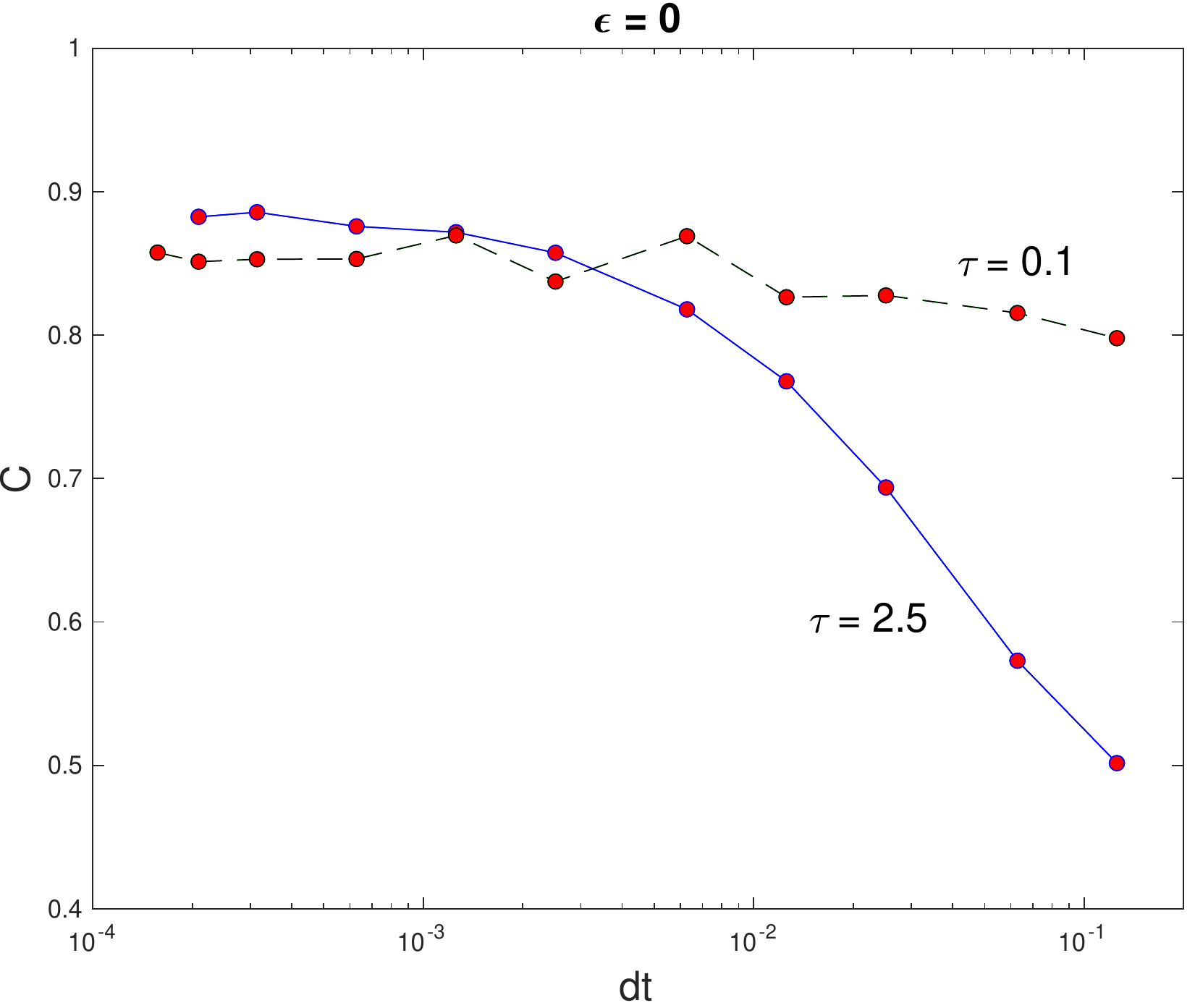}
    \caption{Convergence tests in time step for several set-ups spanning dilute and cold, dense and hot, etc.}
    \label{convergence1}
\end{figure}

\begin{figure}
    \centering
    \includegraphics[width=0.8\columnwidth]{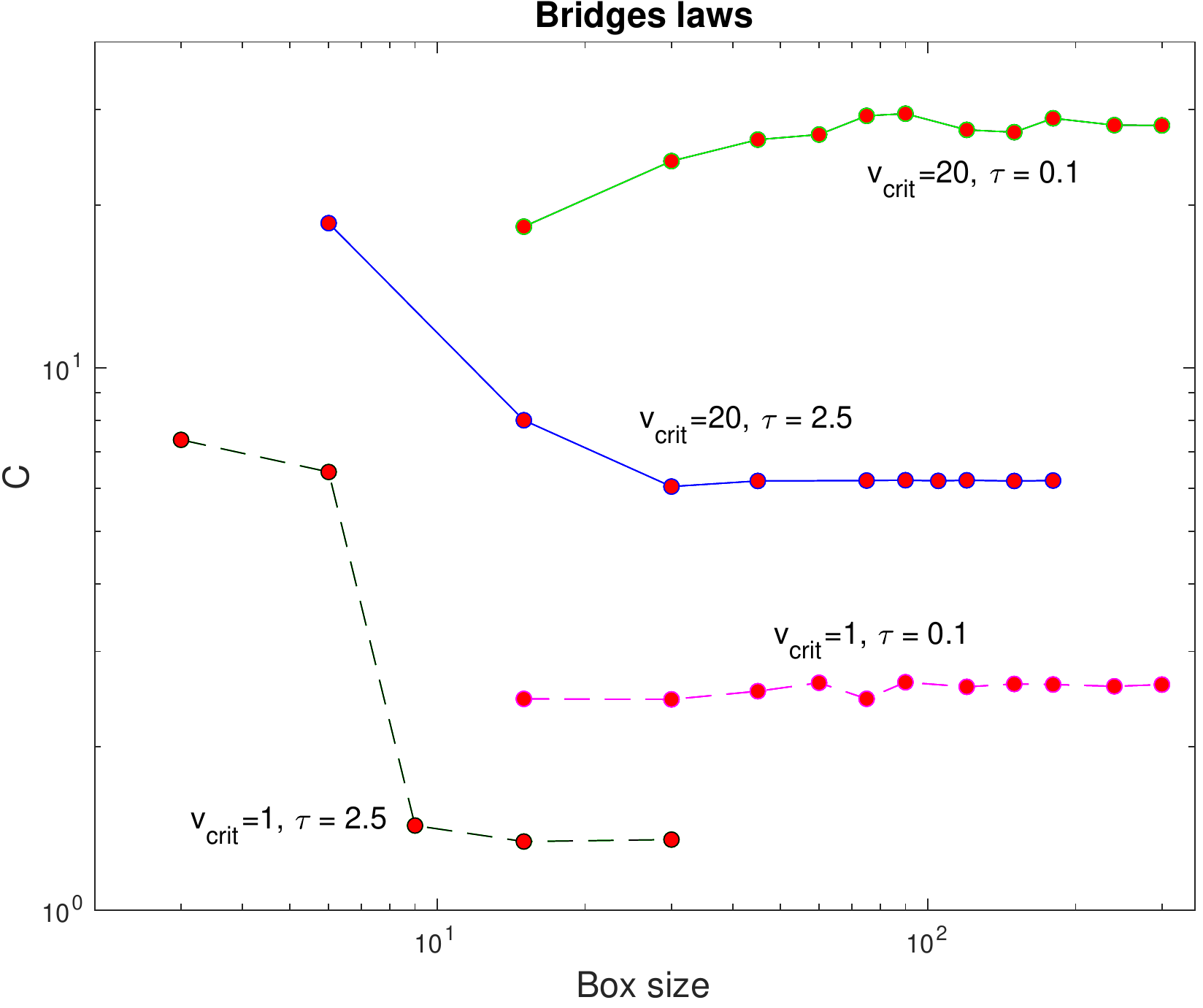}
    \includegraphics[width=0.8\columnwidth]{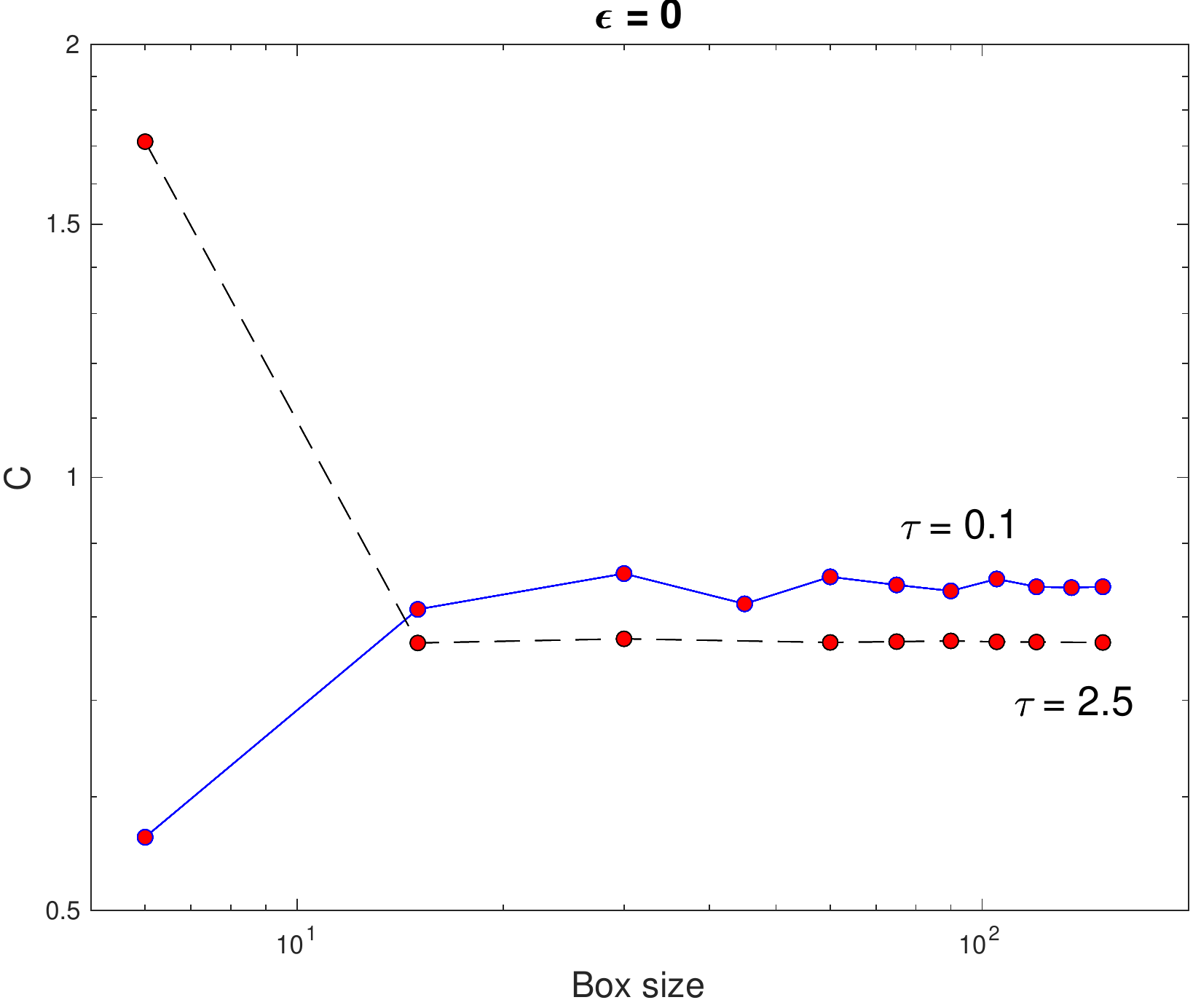}
    \caption{Convergence tests in box size for several set-ups spanning dilute and cold, dense and hot, etc.}
    \label{convergence2}
\end{figure}

Our convergence results are plotted in Figs \ref{convergence1} and \ref{convergence2}, the former showing the velocity dispersion $c$ as a function of $dt$, the latter $c$ as a function of box size. Time steps of $10^{-2}$ or less and a box size of 30 or greater appear to be sufficient in most cases. In our main equilibrium runs in Section 3, we use $dt=10^{-3}$ and a box size of 100. 

\section{Illustrative toy fronts}

In this appendix we calculate thermal fronts using the simple continuum model of Section 5.1.2 with prescribed functions for $\Lambda$ and $k$. Noting the bistability at low $\tau$, we adopt a logistic reaction term and a linear diffusivity, which in suitable units take the form
$$\Lambda= (E-E_C)(E-E_I)(E-E_H), \qquad k = \alpha E, $$
where $E_C< E_I<E_H$ are constant parameters denoting the cold, intermediate, and hot states (respectively), and $\alpha$ is an additional constant. Both $E_C$ and $E_H$ are thermally stable, but $E_I$ is unstable. The basins of attraction of $E_C$ and $E_H$, however, are controlled by their proximity to $E_I$.

These functional choices simplify the integrals in the numerator of \eqref{vf}. The integral of $\Lambda$ becomes simply $-(E_C - E_H)^3(E_C - 2 E_I + E_H)/12$, and is negative when the intermediate state is less than the arithmetic mean of the hot and cold states, $E_I< (E_C+E_H)/2$, and positive otherwise. In other words, the front will tend to move into the cold state when the intermediate state is closer to the cold state, i.e. when its basin of attraction is smaller. Similarly, the front will tend to move into the hot state when $E_I$ is closer to $E_H$. If the three thermal states are equidistant and $k$ is a constant, then $c=0$ and the front profile can be expressed in terms of elliptic integrals.

The second term in \eqref{vf} cannot be evaluated without knowledge of the front profile. It nonetheless simplifies to $-\alpha\int_{-\infty}^\infty (dE/d\xi)^3 d\xi$, which is clearly negative for monotonic front profiles. Thus the linear $k$ law favours the front's movement into the cold state, as discussed in Section 5.1.2. 

\begin{figure}
    \centering
    \includegraphics[width=\columnwidth]{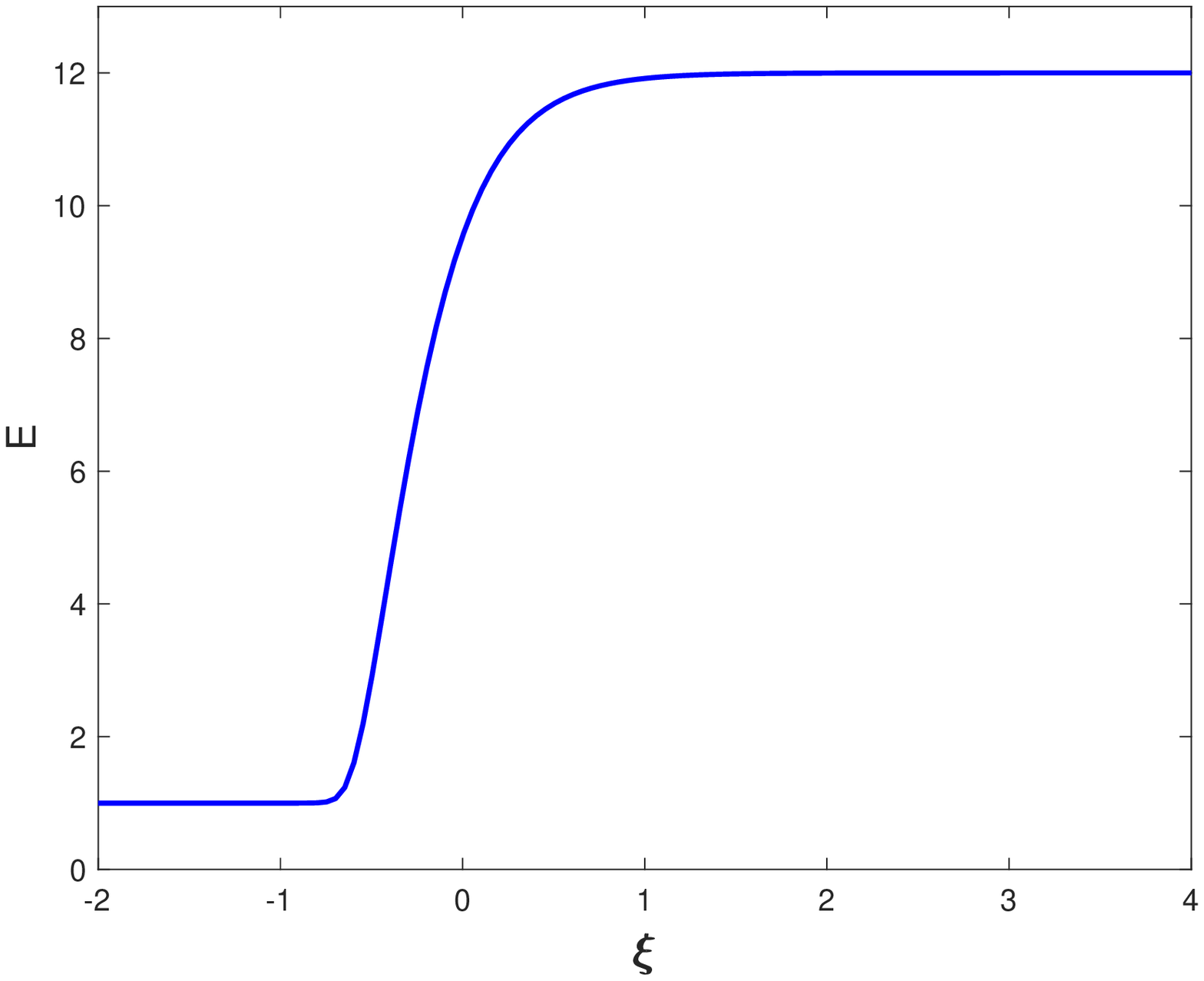}
    \caption{Illustrative front calculated numerically, with parameters $E_C=1$, $E_I=1.5$, $E_H=12$, and $\alpha=0.5$. The front moves to the left into the cold state with a speed $c=-12.3728$.}
    \label{toyfront}
\end{figure}

Finally, we numerically solved \eqref{stefan} using a relaxation method (Press et al.~1986), due to the problem's characteristic stiffness. We plot a representative front solution in Fig.~\ref{toyfront}. As in Fig.~10, the front is sharp at the cold transition, where $k$ is small, and diffuse at the hot transition, where it is an order of magnitude larger.

\end{document}